\documentclass{ptephy_v1}

\preprintnumber{1602.08692} 

\usepackage{color}
\usepackage{graphicx}
\usepackage{amsmath}
\usepackage{color}
\usepackage{bm}

\allowdisplaybreaks[3]

\usepackage[normalem]{ulem}  
\renewcommand\sout{\bgroup \color{blue} \ULdepth=-.5ex \ULset}

\begin{document}

\title{Effective field theory and the scattering process for magnons in the ferromagnet, antiferromagnet, and  ferrimagnet}


\author{Shinya~Gongyo}
\affil{Yukawa Institute for Theoretical Physics, Kyoto University, Kyoto 606-8502, Japan \email{shinya.gongyo@yukawa.kyoto-u.ac.jp}
}
\affil{Theoretical Research Division, Nishina Center, RIKEN, Saitama 351-0198, Japan}
\affil{CNRS, Laboratoire de Math\'ematiques et Physique Th\'eorique, Universit\'e de Tours, 37200 France}
\author{Yuta~Kikuchi}
\affil{Department of Physics, Kyoto University, Kyoto 606-8502, Japan}
\affil{Department of Physics and Astronomy, Stony Brook University, Stony Brook, New York 11794-3800, USA}
\author{Tetsuo~Hyodo}
\affil{Yukawa Institute for Theoretical Physics, Kyoto University, Kyoto 606-8502, Japan}
\author{Teiji~Kunihiro}
\affil{Department of Physics, Kyoto University, Kyoto 606-8502, Japan}

\begin{abstract}
We discuss that a low-energy effective Lagrangian relying on SO(3) $\rightarrow$ SO(2) is applicable for a ferrimagnet as well as a ferromagnet and an antiferromagnet.  
 We derive the Hamiltonian from the Lagrangian using the Poisson and Dirac brackets, 
and determine the particle states. 
The analysis of the particle states shows that 
there exist not only massless modes with the dispersion relations $\omega \propto |\bm{k}|,\, |\bm{k}|^2$, i.e., 
the so-called type-I and type-II Nambu-Goldstone modes, respectively,
but also gapped modes with $\omega \propto m^2+|\bm{k}|^2$. 
We clarify how the coefficients of the terms with one time derivative and those with two time derivatives in the effective Lagrangian 
determine the order parameters specifying whether the system is in a ferromagnetic, antiferromagnetic or ferrimagnetic state;
we stress that the gapped mode resulting form the spontaneous symmetry breaking 
appears only in the ferrimagnetic system and not in the ferromagnetic and antiferromagnetic systems. 
We also establish the power counting scheme and calculate the scattering amplitudes and thereby the scattering lengths 
between the two Nambu-Goldstone bosons.  
We show that the scattering length of the gapped mode is finite
  and proportional to the gap. This characteristic property of the
  gapped NG mode can be used to discriminate it from gapped
  excitations which originate in other mechanisms. Finally, we study the effects of the explicit symmetry breaking that are given by an external magnetic field and a single-ion anisotropy, and show that the external magnetic fields do not have any effects 
on the scattering amplitudes in all the spin systems as was  known for the ferromagnet system. In contrast, the anisotropy does affect the scattering amplitudes, 
the phase shift, and the scattering length except for spin 1/2 systems. 
This result supports the possibility of the Efimov effect in spin systems discussed in previous studies.
\end{abstract}

\subjectindex{B31, B36, I46}

\maketitle

\section{Introduction}
The method of effective field theory (EFT) is a powerful and successful tool to understand the low-energy behaviors  
in various systems extending from particle physics to condensed matter physics 
\cite{Weinberg:1978kz,Gasser:1983yg,Gasser:1984gg,Leutwyler:1993gf,Burgess:1998ku,Brauner:2010wm,Scherer:2012xha}.
 The EFT is based on the spontaneous symmetry breaking pattern in the underlying theory, 
and is constructed in terms of the corresponding Nambu-Goldstone (NG) fields, which leads
 to the systematic and model-independent analysis of the low-energy behavior. 
The method has been originally employed in the field of particle physics.
 More specifically, 
chiral perturbation theory, which is an EFT relying on chiral symmetry breaking 
and the NG modes, 
i.e. pions, has succeeded in describing the low-energy behaviors in quantum chromodynamics (QCD)
 such as the pion scattering processes and hadron structures \cite{Gasser:1983yg,Gasser:1984gg,Scherer:2012xha}. 
Moreover, the EFT has been used for the analysis of the hadron properties at finite-density QCD, 
in various phases including the chiral-symmetry broken or partially restored phases, 
 the kaon-condensed phase, the color superconductivity phase and so on.
 Recently, an EFT has been developed to incorporate spatial symmetry breaking
 \cite{Lee:2015bva,Hidaka:2015xza}
for describing the inhomogeneous chiral phases, 
which may appear in high-density quark matter.

The EFT approach has also been widely applied to condensed matter physics
\cite{Hasenfratz:1993vf,Leutwyler:1993gf,Leutwyler:1996er,roman9709298effective,Hofmann:1998pp,Burgess:1998ku,Hofmann:2001ck,Brauner:2010wm}
: For instance, the translational and rotational symmetries are
spontaneously broken in the 
ground state of the solid states, and the spin rotational symmetry is also  broken in spin systems with some order. 
Then according to the symmetry breaking patterns, 
the EFTs have been constructed 
in terms of  spin waves, i.e. magnons, as the NG modes in spin systems 
\cite{Leutwyler:1993gf,roman9709298effective} and the sound wave, 
i.e. phonon, in solid-state systems \cite{Leutwyler:1996er}. 
In particular, using the EFT for spin systems enables us to analyze the low-energy properties 
such as the nature of the low-energy modes i.e., magnons 
and their scattering processes 
in ferromagnets and antiferromagnets \cite{Hofmann:1998pp}: 
It was shown that the difference in the order parameters between ferromagnet and antiferromagnet leads 
to different characteristics of the low-energy (NG) modes 
and amplitudes of the  magnon-magnon scattering even if the patterns of the symmetry breakings are the same.

In this paper, we develop a unified EFT for the collinear phases
of spin systems which include ferromagnets, antiferromagnets and ferrimagnets
where the spin directions at lattice sites are all aligned or anti-aligned.
On the basis of this EFT, 
we elucidate the characteristics of the low-energy 
magnon-magnon scattering, which should be of basic importance in examining the possible Efimov effect of magnons as the NG modes in all the spin-ordered phases.
We remark that Efimov effects of NG bosons have been discussed for magnons in the ferromagnet
 \cite{Nishida:2012hf,nishida2013electron} and pions in QCD \cite{Hyodo:2013zxa}. 

The EFT approach has also been utilized to count the number of NG modes 
as well as to understand the dispersion relations. In systems without Lorentz invariance including finite density QCD, spin systems, 
and solid systems the number of NG modes can be less than that of the broken generators, 
while in Lorentz-invariant systems, the number of NG modes coincides with 
that of the broken generators \cite{Nielsen:1975hm, Miransky:2001tw, Schafer:2001bq, Nambu:2004yia, Hidaka:2012ym,Watanabe:2012hr}. 
The NG modes in systems without Lorentz invariance have been classified into type-I and type-II, 
whose energies are proportional to odd and even powers of the momentum, respectively, 
while in Lorentz-invariant systems, the NG modes always have a linear dispersion relation. 
In fact, in the ferromagnet, the number of NG modes is less than that of the broken generators, 
and the NG mode shows the quadratic dispersion relation. On the other hand, in the antiferromagnet, 
the number of NG modes coincides with that of the broken generators, 
and the NG modes show the linear dispersion relation.

Quite recently, it has been clarified on the basis of the EFT that gapped modes 
may also appear in addition to the type-II NG modes as a result of the spontaneous symmetry breaking
 \cite{Kapustin:2012cr, Nicolis:2012vf, Gongyo:2014sra, Hayata:2014yga,Beekman:2014cba}.
Therefore the NG modes can be classified into at least three types, i.e., type-I, type-II and gapped modes.  
The gapped NG modes are accompanied by the type-II modes, so that a reduction of the number of NG modes does not occur \cite{Gongyo:2014sra}. 
A general relation between the appearance of gapped NG modes 
and some specific nature of the order parameters has been 
discussed \cite{Hayata:2014yga,Beekman:2014cba}. 
In this paper, we discuss that the gapped NG mode is identified with the magnon in ferrimagnets from the perspective of the order parameters and dispersion relations. We make clear how the EFT works by constructing the power counting scheme and that the gapped mode is reliable and affects the magnon-magnon scattering in ferrimagnets. We find nonzero scattering lengths only for processes involving the gapped NG mode. The strength of the scattering length is given by the gap of the mode multiplied by a universal factor.

\section{Effective field theory for SO(3) $\rightarrow$ SO(2) and the scattering amplitude}
\label{EFT}
\subsection{Construction of the effective Lagrangian for SO(3) $\rightarrow$ SO(2)}

We consider the spin systems described by the Hamiltonian which is invariant under the global SO(3) transformation of spins $S_m^a~(a=1,2,3)$:
\begin{align}
\mathcal{H}
&=\mathcal{H}_{\mathrm{inv}}( S_m^a ) , \\
S_m^a &\rightarrow g^{ab}S^b_m, ~~~g^{ab} \in \mathrm{SO}\left(3\right) .
\label{spin}
\end{align}
We assume that the ground state is in the collinear phases 
where all of the spins defined on each lattice site $m$ are aligned or anti-aligned.
 When we set the 3-axis in the direction of the spin alignment and anti-alignment, the order parameter for the alignment is given by the magnetization, $\left<\sum _m S_m^3\right>$ while that for the anti-alignment is given by the staggered magnetization, $\left<\sum _m (-1)^m S_m^3\right>$. When the order parameter is nonzero, the global SO(3) symmetry is spontaneously broken down to SO(2) corresponding to the rotation around the 3-axis.

As a concrete example, let us consider the Heisenberg model in a ferromagnet system,
\begin{align}
\mathcal{H}= -J\sum_{\left<m,n\right>} S_m^a S_n^a, 
\label{eq:Heisenberg}
\end{align}
where $\left<m,n \right>$ denotes the summation over the nearest neighbors. The Hamiltonian is invariant under the global SO(3) transformation~\eqref{spin}.
The Hamiltonian also has spatial translational and rotational symmetry in the continuum limit. 
However, in the ground state, all of the spins are aligned in one direction, and the spin rotational symmetry SO(3) breaks down to SO(2) spontaneously. We emphasize that the following effective field theory (EFT) is applicable to any underlying theory with the symmetry breaking pattern being SO(3)$\to $SO(2). The EFT describes the low-energy behavior model-independently, just using the information on a symmetry breaking pattern and the order parameter.

The construction of EFT relying on a general symmetry breaking
  pattern has been discussed in terms of the Maurer-Cartan one form
  \cite{Andersen:2014ywa,Watanabe:2014fva}. Here we briefly review this procedure to give a basis of the main results of this paper. Motivated by the symmetry breaking pattern of the collinear phases, SO(3) $\rightarrow$ SO(2), we focus on the low-energy EFT using the Maurer-Cartan one form \cite{Coleman:1969sm, Callan:1969sn}:
\begin{align}
\alpha_\mu \left(\pi \right)=-i U^{-1}(\pi)\partial_\mu U(\pi),
\end{align}
with $\partial _\mu= \left( \frac{\partial}{\partial t}, \frac{\partial}{\partial x_i} \right)~\left(\mu =0,\dots,3, i=1,\dots, 3 \right)$ and the representative of the coset space SO(3)/SO(2) being $U(\pi)$. The representative $U(\pi)$ is characterized by two NG fields $\pi ^\alpha \left(\alpha =1,2 \right)$, corresponding to the number of broken generators, dim$\left(\mathrm{SO}(3)/\mathrm{SO}(2)\right)=2$. Under the SO(3) global transformation $g \in$ SO(3), $U(\pi)$ is transformed into 
\begin{align}
U(\pi) \rightarrow gU(\pi)h^{-1}\left(\pi, g\right) ,\label{transformation_rule}
\end{align} 
with $h\left(\pi, g\right) \in$ SO(2). Although in this paper, we focus on the system in (1+3) dimensions, it is easy to generalize the analysis to other dimensions.

The transformation behavior of $\alpha_\mu \left( \pi \right)$ is given by
\begin{align}
\alpha_\mu \left( \pi \right) &\rightarrow h\left(\pi, g\right) \alpha_\mu \left(\pi \right)h^{-1}\left(\pi, g\right) \nonumber \\
&\quad +i^{-1}h\left(\pi, g\right) \partial_\mu h^{-1}\left(\pi, g\right).
\end{align}
Using the derivative expansion appropriate to discuss the low-energy behavior, we construct the effective Lagrangian 
as is done in chiral perturbation theory \cite{Gasser:1983yg,Gasser:1984gg}. 
Note that the system has spatial translational and rotational symmetry but not Lorentz symmetry, 
as in the nonrelativistic system.\footnote{
 In Refs.~\cite{Hasenfratz:1993vf,roman9709298effective,Hofmann:2001ck}, 
the effect of the microscopic structure of the underlying theory that breaks translational and rotational symmetry enters in the higher order 
terms of the derivative expansion. Because here we concentrate on the leading order calculation, 
the EFT can be constructed under the invariance of spatial translational and rotational symmetry.} 
According to a previous study \cite{Leutwyler:1993gf}, the effective Lagrangian 
may possess invariant terms up to the total derivative, which can lead to condensates of the charge densities; 
as also discussed in Sec.\ref{condensates_and_coefficients}. 
To construct the Lagrangian explicitly, we make the following decomposition:
\begin{align}
\alpha_{\mu \parallel}(\pi)&\equiv \frac{1}{2}\mathrm{Tr}(T^3 \alpha_\mu (\pi)) T^3 ,\\
\alpha_{\mu \perp}(\pi)&\equiv \frac{1}{2}\mathrm{Tr}(T^\alpha \alpha_\mu (\pi)) T^\alpha ,
\end{align}
with  $T^\alpha (\alpha=1,2)$ and $T^3$ broken and unbroken generators, respectively,
\begin{align}
\left[T^a\right]_{bc} = -i \epsilon_{abc}~~\left(a=1,2,3\right) .
\end{align}
The decomposition leads to the transformation rule of $\alpha_{\mu \parallel}(\pi)$ and $\alpha_{\mu \perp}(\pi)$,
\begin{align}
\alpha_{\mu \parallel}(\pi)&\rightarrow  h(\pi,g)\alpha_{\mu \parallel}(\pi)h^{-1}(\pi,g) \nonumber \\
&\quad +i^{-1}h(\pi,g) \partial_\mu h^{-1}(\pi,g),\\
\alpha_{\mu \perp}(\pi)&\rightarrow h(\pi,g)\alpha _{\mu \perp}(\pi)h^{-1}(\pi,g).
\end{align}
Using the decomposition, the invariant effective Lagrangian reads up to $O\left(\partial_0 ^2\right)$ and $O\left(\partial_i ^2\right)$ orders, 
\begin{align}
\mathcal{L} 
   &=-\frac{\Sigma}{2}\mathrm{Tr}\left[T^3 \alpha_{0 \parallel}\right]+ \frac{F_t^2}{4}\mathrm{Tr}\left[\alpha_{0 \perp}\right]^2
   \notag \\
   &\quad -\frac{F^2}{4}\mathrm{Tr}\left[\alpha_{i \perp}\right]^2 +O(\partial _0^3, \partial_i^4)\notag \\
&= -\frac{\Sigma}{2}\mathrm{Tr}\left[ T^3\alpha_0\right] 
+\frac{F_t^2}{8}\mathrm{Tr}\left[T^\alpha \alpha_0\right] \mathrm{Tr}\left[T^\alpha \alpha_0\right] \notag \\
   &\quad -\frac{F^2}{8}\mathrm{Tr}\left[T^\alpha \alpha_i\right] \mathrm{Tr}\left[T^\alpha \alpha_i\right] +O(\partial _0^3, \partial_i^4) \notag \\
&= \frac{i\Sigma}{2}\mathrm{Tr}\left[ T^3U^{-1} \partial_0U\right] 
\notag \\
   &\quad -\frac{F_t^2}{8}\mathrm{Tr}\left[T^\alpha U^{-1} \partial_0 U\right] \mathrm{Tr}\left[T^\alpha U^{-1}\partial_0 U\right]  \notag \\
&\quad +\frac{F^2}{8}\mathrm{Tr}\left[T^\alpha U^{-1} \partial_i U\right] \mathrm{Tr}\left[T^\alpha U^{-1} \partial_i U\right] +O(\partial _0^3, \partial_i^4) , \label{eff_L}
\end{align}
with three parameters, $\Sigma, F_t,$ and $F$. The first term is invariant only up to the total derivative. We take $\hbar=1$ and $c\neq 1$, and the dimensions of the three parameters are given as follows: $[\Sigma]=L^{-3}, [F_t^2]=T\cdot L^{-3}, [F^2]=T^{-1}\cdot L^{-1}$. The relation between the coefficients and the order parameter is discussed in Sec.\ref{condensates_and_coefficients}. 

By using the coordinates of the coset space, $U=e^{i\pi^\alpha T^\alpha/F}$ with the NG fields $\pi ^\alpha$, and expanding the exponential, the Lagrangian reduces to
\begin{align}
\mathcal{L} &\equiv \mathcal{L}^{(1,0)}+\mathcal{L}^{(2,0)}+\mathcal{L}^{(0,2)}+\dotsb,  \label{L_pi}\\ 
\mathcal{L}^{(1,0)}&=-\frac{\Sigma}{2F^2}\epsilon ^{\alpha \beta}\pi^\alpha \partial_0 \pi ^\beta 
+\frac{\Sigma}{24F^4}\epsilon^{\alpha \beta}\pi ^\alpha \partial_0 \pi ^\beta \pi^\gamma \pi^\gamma+\dotsb, \\
\mathcal{L}^{(2,0)}
&=\frac{1}{2v^2}\partial _0 \pi^ \alpha \partial _0 \pi^\alpha \nonumber \\
&\quad -\frac{1}{6v^2F^2}\left[\partial_0 \pi^\alpha \partial_0 \pi^\alpha \pi^\beta \pi^\beta-\pi^\alpha\partial_0\pi^\alpha \pi^\beta \partial_0 \pi^\beta\right]+\dotsb ,  \\
\mathcal{L}^{(0,2)}
&=-\frac{1}{2}\partial _i \pi^ \alpha \partial _i \pi^\alpha \nonumber \\
&\quad +\frac{1}{6F^2}\left[\partial_i \pi^\alpha \partial_i \pi^\alpha \pi^\beta \pi^\beta-\pi^\alpha\partial_i\pi^\alpha \pi^\beta \partial_i \pi^\beta\right]+\dotsb , \label{L02}
\end{align}
with $v\equiv F/F_t$. Note that the terms with odd-power of $\pi$ fields do not appear 
in the Lagrangian because of the invariance under the ``parity'' transformation,
\begin{align}
h_p U\left(\pi \right)h^{-1}_p=U\left(-\pi \right),~ h_p = \begin{pmatrix} -1 &~&~ \\ ~& -1 &~ \\ ~&~&1 \end{pmatrix},
\end{align}
with $h_p \in$ SO(2). The ``parity'' invariance is guaranteed by the fact that the coset space is a symmetric space, as a consequence of $[\mathcal{G}-\mathcal{H},\mathcal{G}-\mathcal{H}]\subset\mathcal{H}$.

\subsection{Dispersion relation and power counting}
\label{Dispersion relation and power counting}

As we shall show in Sec.\ref{condensates_and_coefficients}, 
the nonzero values of $1/v$ and $\Sigma$ are the consequence of the existence 
of the corresponding order parameters. Let us consider the following three cases
separately: 
(a) $1/v=0, \Sigma \ne 0$, (b) $1/v\ne 0, \Sigma = 0$, and (c) $1/v\ne0, \Sigma \ne 0$. 
In the case of (a), the dispersion relation of the NG mode derived from the equation of motion is given by
\begin{align}
\omega=\frac{F^2}{\Sigma} {\bm k }^2 , \label{dispersion_typeII}
\end{align}
and the number of degrees of freedom (DOF) of the mode is one. 
In the case of (b), the dispersion relation is given by $\omega=v |{\bm k }|$, 
which is similar to that of pions in QCD, and the number of DOF is two. 
In the case of (c), the dispersion relations are given by
\begin{align}
\omega^{II}&\equiv \frac{v^2\Sigma}{2F^2}\left[-1+\sqrt{1+\left(\frac{2F^2}{v\Sigma}\right)^2\bm{k}^2}\right] \simeq \frac{F^2}{\Sigma}\bm{k}^2 , \notag \\
\omega^{M}&\equiv \frac{v^2\Sigma}{2F^2}\left[1+\sqrt{1+\left(\frac{2F^2}{v\Sigma}\right)^2\bm{k}^2}\right] \simeq \frac{\Sigma}{F_t^2}+\frac{F^2}{\Sigma}\bm{k}^2, \label{dispersions}
\end{align}
and the number of DOF is two \cite{Kapustin:2012cr,Watanabe:2014fva,Gongyo:2014sra,Miransky:2001tw,Schafer:2001bq}.  
In fact, these dispersion relations
 coincide with the behaviors of the long wavelength modes obtained by the Heisenberg models 
in a ferromagnet, an antiferromagnet, and a ferrimagnet, respectively, 
where the Holstein-Primakoff transformation is utilized \cite{kittel1963quantum,brehmer1997low,pati1997density}.
Thus we stress that the EFT describes not only the low-energy behavior of the ferromagnet and
  antiferromagnet \cite{Leutwyler:1993gf, Hofmann:1998pp}, but also that of the ferrimagnet. 

Note that in the case of (c), the dispersion relation of 
one of the NG modes tends to coincide with that of the NG mode 
in the case of (a), i.e., the magnon in a ferromagnet, whereas the other tends to be a gapped mode 
with the gap energy $\nu_{M}=\Sigma/F_{t}^{2}$ in the low energy regime. The same dispersion relation has been discussed in the context of the kaon-condensed phase for finite density QCD \cite{Miransky:2001tw,Schafer:2001bq}. The gapped mode at finite density comes from the spontaneous breaking of a symmetry generated by the charge coupled to the chemical potential in an underlying Hamiltonian \cite{Nicolis:2012vf}. The gapped mechanism also occurs if a symmetry is both spontaneously and explicitly broken and if the broken charge couples to the external field \cite{Watanabe:2013uya}. In contrast, the gapped mode in the ferrimagnet appears without the chemical potential or the external field. The gapped mode becomes the type-I NG mode in the limit of $\Sigma \to 0$, which also indicates that the gapped mode is related to the spontaneous symmetry breaking.

It should be instructive to note that
 the free part of the effective Lagrangian for the ferromagnet 
is analogous to the classical Lagrangian of a fast rotating rigid pendulum 
under the gravitational force, and the dispersion relation of the type-II NG mode, 
Eq.(\ref{dispersion_typeII}), is identified with the frequency of the precession motion 
by regarding the gravitational constant $g\sim \bm{k}^{2}$.\footnote{Y.Hidaka, informal lecture 
at Kyoto University.} When the angular velocity of the pendulum decreases, 
nutation is visible in addition to the precession motion \cite{landau1960classical}. 
In terms of the Lagrangian, the nutation motion can be described by including the higher 
order time derivatives of the rigid-body coordinate, as in the case of the ferrimagnet. 
The dispersion relation of the gapped mode is then identified with the frequency 
of the nutation, which remains finite in the $g\to 0$ limit. 

Once the dispersion relation is determined, we can establish the power counting 
scheme\cite{roman9709298effective}. The underlying theory contains the natural scales 
of the energy and momentum. We specify the energy scale $\nu$ by the first gapped excitation 
whose origin is not related to the spontaneous symmetry breaking. 
The momentum scale $\Lambda$ is determined by the inverse of the lattice spacing 
in the underlying theory $\sim 1/a$. Generically, the EFT description is based 
on the double expansion of $\partial_{0}/\nu$ and $\partial_{i}/\Lambda$. 
The temporal and spatial components are related to each other through the dispersion relation. 
We define the small dimensionless parameter $p$ as
\begin{align}
\begin{cases}
\frac{\partial_{0}}{\nu} \sim (\frac{\partial_{i}}{\Lambda})^{2} 
\sim p^{2} & 1/v= 0, \Sigma\neq 0 \text{ case (a)}\\
\frac{\partial_{0}}{\nu} \sim\frac{\partial_{i}}{\Lambda} \sim p & 1/v\neq 0, \Sigma=0 \text{ case (b)} \\
\frac{\partial_{0}}{\nu}\sim \frac{\nu_{M}}{\nu} \sim  (\frac{\partial_{i}}{\Lambda})^{2}
\sim p^{2} & 1/v\neq  0, \Sigma\neq 0 \text{ case (c)}
\end{cases} ,
\end{align}
and sort out the effective Lagrangian in powers of $p\ll 1$. The low energy phenomena can be calculated with the terms with small powers of $p$. We note that the gapped NG mode in Eq.~\eqref{dispersions} is meaningful only when $\nu_{M}\ll \nu$.

\subsection{Construction of the Hamiltonian and particle states}
\label{Construction of the Hamiltonian and particle states}

To calculate the scattering amplitudes, we construct the asymptotic states. 
In this section, we derive the Hamiltonian from the free part of the Lagrangian, 
and define the particle states. The Lagrangian at $\mathcal{O}(p^{2})$ in Eq.~(\ref{L_pi}) is given by
\begin{align}
\mathcal{L}_0=-\frac{\Sigma}{2F^2}\epsilon ^{\alpha \beta}\pi^\alpha \partial_0 \pi ^\beta 
+\frac{1}{2v^2}\partial _0 \pi^ \alpha \partial _0 \pi^\alpha-\frac{1}{2}\partial _i \pi^ \alpha \partial _i \pi^\alpha.
\end{align}
How to construct the Hamiltonian depends on whether $1/v$ is zero or not, because if it is zero, 
the EFT must be formulated as that for a constrained system \cite{Gongyo:2014sra}. 
Thus we consider two cases: (i) $1/v=0$ (ferromagnet case) and (ii) $1/v \ne 0$ (antiferromagnet and ferrimagnet cases). According to the Dirac-Bergmann method \cite{Dirac:1950pj,Bergmann:1949zz}, the construction of the (total) Hamiltonian in the EFT has been generally discussed \cite{Gongyo:2014sra}. In this paper, we focus on the construction of the Hamiltonian in the spin systems.

(i) The canonical momenta are given by
\begin{align}
P^{\alpha} &\equiv \frac{\partial \mathcal{L}}{\partial \dot{\pi} ^\alpha} 
=\frac{\Sigma}{2F^2}\epsilon^{\alpha\beta}\pi ^\beta.
\end{align}
The canonical momenta are not expressed with $\dot{\pi}^\alpha$, and thus primary constraints $\phi^\alpha$ appear,
\begin{align}
\phi^\alpha\equiv P^\alpha -\frac{\Sigma}{2F^2}\epsilon^{\alpha\beta}\pi ^\beta.
\end{align}
The constraints $\phi ^\alpha$ are second class,
\begin{align}
C^{\alpha \beta} \left(\bm{x},\bm{y} \right)\equiv \left\{\phi^\alpha(\bm{x}),\phi^\beta(\bm{y}) \right\}=-\frac{\Sigma}{F^2}\epsilon^{\alpha \beta}\delta^3(\bm{x}-\bm{y})
,
\end{align}
with $\left\{\dots \right\}$ being the Poisson bracket.
The total Hamiltonian $\mathcal{H}^T_0 $ is given by
\begin{align}
\mathcal{H}^T_0 &\equiv \dot{\pi}^\alpha P^\alpha -\mathcal{L}_0 +\lambda^\alpha \phi^\alpha \notag \\
&=\frac{1}{2}\partial_i \pi ^\alpha \partial_i \pi ^\alpha + \lambda^\alpha \phi ^\alpha \notag \\
&\equiv \mathcal{H}_0 + \lambda^\alpha \phi ^\alpha,
\end{align}
where $\lambda ^\alpha$ are Lagrange multipliers, and $\mathcal{H}_0$ is the canonical Hamiltonian. In the second line, $\dot{\pi}^\alpha \phi ^\alpha$ has been absorbed into the last term $\lambda ^\alpha \phi ^\alpha$. The Lagrange multipliers $\lambda ^\alpha$ are given by
\begin{align}
\lambda^\alpha =-\frac{F^2}{\Sigma}\epsilon ^{\alpha\beta}\partial_i^2 \pi ^\beta,
\end{align}
using the consistent condition of the constraints for the time evolution,
\begin{align}
\left\{ \phi ^\alpha, \int \mathcal{H}^T_0 d^3y \right\}
&=\left\{ \phi ^\alpha, \int \left(\frac{1}{2}\partial_i \pi ^\alpha \partial_i \pi ^\alpha + \lambda^\alpha \phi ^\alpha \right)d^3y \right\} \nonumber \\
&\approx \partial_i ^2 \pi ^\alpha - \lambda^\beta \frac{\Sigma}{F^2}\epsilon^{\alpha \beta} 
\approx 0,
\end{align}
where ``$\approx$" is defined as an equality under the constraint conditions. Thus the total Hamiltonian reduces to
\begin{align}
\mathcal{H}^T_0 
&=\frac{1}{2}\partial_i \pi ^\alpha \partial_i \pi ^\alpha - \frac{F^2}{\Sigma}\epsilon ^{\alpha\beta}\partial_i^2 \pi ^\beta\left( P^\alpha -\frac{\Sigma}{2F^2}\epsilon^{\alpha\gamma}\pi ^\gamma\right) \nonumber \\
&=-\frac{F^2}{\Sigma}\epsilon ^{\alpha\beta}P^\alpha\partial_i^2 \pi ^\beta. 
\end{align}
The total Hamiltonian leads to the same equation of motion for the NG fields obtained from the Lagrangian $\mathcal{L}_0$.

The quantization of the constraint system is performed by replacing the Dirac bracket with the commutation relation,
\begin{align}
\left\{F\left( {\bm x}\right), G\left( {\bm y}\right)\right\}_D \rightarrow \frac{1}{i} \left[F\left( {\bm x}\right), G\left( {\bm y}\right)\right],
\end{align}
for arbitrary functions $F\left( {\bm x}\right)$ and $G\left( {\bm y}\right)$.
 Instead of the total Hamiltonian $\mathcal{H}^T_0$ and the Poisson bracket, we use the canonical Hamiltonian $\mathcal{H}_0$ and the Dirac bracket defined by
\begin{align}
\left\{F\left( {\bm x}\right), G\left( {\bm y}\right) \right\} _{D}
&\equiv \left\{F\left( {\bm x}\right), G\left( {\bm y}\right) \right\} \nonumber \\
&\quad -\int d^3 z d^3 w\left\{ F\left(  {\bm x}\right), \phi ^\beta\left( {\bm w}\right) \right\}\nonumber \\
&\quad\times \left(C^{-1}\right)^{ \beta \gamma}\left( {\bm w}, {\bm z}\right)\left\{ \phi^\gamma\left(  {\bm z}\right), G\left(  {\bm y}\right) \right\}. 
\end{align}

The commutation relations between the NG fields and the momenta are given as follows:
\begin{align}
[\pi ^\alpha\left( {\bm x}\right), \pi ^\beta \left( {\bm y}\right)]&= i\epsilon ^{\alpha \beta} \frac{F^2}{\Sigma}\delta ^3(\bm{x}-\bm{y}), \label{com_pp} \\
[\pi ^\alpha \left( {\bm x}\right), P ^\beta\left( {\bm y}\right)]&=\frac{i}{2}\delta ^{\alpha\beta}\delta ^3(\bm{x}-\bm{y}), \label{com_pP} \\
[P ^\alpha \left( {\bm x}\right), P ^\beta \left( {\bm y}\right)]&=\frac{i\Sigma}{4F^2}\epsilon^{\alpha \beta}
\delta^3(\bm{x}-\bm{y}) . \label{com_PP} 
\end{align}
The Heisenberg equation corresponds to the equation of motion obtained by the Lagrangian. 
The NG fields and the momenta are expressed with the creation and annihilation operators,
\begin{align}
\pi ^\alpha(\bm{x},t)&=\frac{F}{\sqrt{\Sigma}}\int \frac{d^3k}{\left(2\pi\right)^{3}}\left\{\epsilon^\alpha a(\bm{k}) e^{-ikx}+\epsilon ^{\ast \alpha} a^{\dagger}(\bm{k}) e^{ikx}\right\},\label{NG_in_fero} \\
P^\alpha (\bm{x},t)&=\frac{i\sqrt{\Sigma}}{2F}\int \frac{d^3k}{\left(2\pi\right)^{3}}\left\{\epsilon^\alpha a(\bm{k}) e^{-ikx}-\epsilon ^{\ast \alpha} a^{\dagger}(\bm{k}) e^{ikx}\right\},
\end{align}
where $(\epsilon^1,\epsilon^2)=(1,-i)/\sqrt{2}$, and $kx=\omega x_0 -{\bm k}\cdot {\bm x}$ with $\omega$ defined in Eq.~(\ref{dispersion_typeII}). 
The creation and annihilation operators satisfy the following commutation relations:
\begin{align}
\left[a(\bm{k}),a^{\dagger}(\bm{q})\right]=&(2\pi)^3 \delta^3(\bm{k}-\bm{q}), \notag \\
\Big[a(\bm{k}),a(\bm{q})\Big]=&
\left[a^{\dagger}(\bm{k}),a^{\dagger}(\bm{q})\right]=0. \label{com_cre}
\end{align}
Note that although the number of NG fields is two, the number of of creation operators (annihilation operators) is one, corresponding to the number of physical degrees of freedom in the constrained system with two NG fields and two constraints, $(2\times2 -2)/2=1$.

Using these operators, we define the vacuum and $n$-particle states:
\begin{align}
a(\bm{k})\left|0\right> &=0, \\
\left|\bm{k}_1,\bm{k}_2,\cdots,\bm{k}_n \right> &\equiv a^\dagger (\bm{k}_1)a^\dagger (\bm{k}_2)\cdots a^\dagger (\bm{k}_n)\left|0\right> 
.
\end{align}

(ii) In the case of $1/v\neq 0$, the corresponding momenta are given by
\begin{align}
P^{\alpha} &\equiv \frac{\partial \mathcal{L}}{\partial \dot{\pi} ^\alpha} 
=\frac{\Sigma}{2F^2}\epsilon^{\alpha\beta}\pi ^\beta +\frac{1}{v^2}\dot{ \pi} ^\alpha.
\end{align}
The $\dot{\pi}^\alpha$ are replaced by $P^\alpha$, and no constraints appear. The Hamiltonian $\mathcal{H}_0$ is obtained by
\begin{align}
\mathcal{H}_0 &\equiv \dot{\pi}^\alpha P^\alpha -\mathcal{L}_0  \notag \\
&=\frac{v^2}{2}\left(P^\alpha -\frac{\Sigma}{2F^2}\epsilon ^{\alpha \beta}\pi^\beta \right)^2 +\frac{1}{2}\partial_i \pi^\alpha \partial_i \pi ^\alpha .
\end{align}
We remark that a general analysis on the relation among
 terms with two time derivatives, constraints, and dispersion relations 
has been given in \cite{Gongyo:2014sra}.

Then we replace the Poisson brackets between the NG fields 
and the corresponding momenta with commutation relations:
\begin{align}
[\pi ^\alpha, \pi ^\beta]&= 0,\notag \\
[\pi ^\alpha, P ^\beta]&=i\delta^{\alpha\beta}\delta^3(\bm{x}-\bm{y}), \notag \\
[P ^\alpha, P ^\beta]&=0.
\end{align}
The Heisenberg equations are now calculated to be
\begin{align}
i\dot{\pi}^\alpha (\bm{x},t) &=\left[\pi ^\alpha (\bm{x},t), \int \mathcal{H}_0 (\bm{y},t)d^3y \right] \notag \\
&=iv^2\left(P^\alpha (\bm{x},t) -\frac{\Sigma}{2F^2}\epsilon^{\alpha \beta}\pi^{\beta } (\bm{x},t)\right), \\
i\dot{P}^\alpha (\bm{x},t) &=\left[P^\alpha (\bm{x},t), \int \mathcal{H}_0 (\bm{y},t)d^3y \right] \notag \\
&=iv^2\frac{\Sigma}{2F^2}\epsilon ^{\beta \alpha} (\bm{x},t)\Biggl[P^\beta (\bm{x},t) \nonumber \\
&\quad -\frac{\Sigma}{2F^2}\epsilon^{\beta \delta}\pi^{\delta } (\bm{x},t)\Biggr] +i\partial_i ^2\pi ^\alpha (\bm{x},t ),
\end{align}
The NG fields and the momenta are expanded in terms of the creation and annihilation operators:
\begin{align}
\pi ^\alpha (\bm{x},t)&= \frac{F}{\sqrt{\Sigma}}\int \frac{d^3k}{\left(2\pi\right)^3\sqrt{\bar{\omega}\left({\bm k}\right)}} \notag \\
&\quad \times \Big\{\epsilon ^\alpha a_{II}(\bm{k})e^{-ik^{II}x}+\epsilon ^{*\alpha}a^\dagger_{II}(\bm{k})e^{ik^{II}x}  \nonumber \\
&\quad +\epsilon ^{* \alpha} a_M(\bm{k})e^{-ik^Mx}+\epsilon ^{\alpha} a^\dagger_M(\bm{k})e^{ik^Mx} \Big\},  \label{NG_in_feri} \\
P ^\alpha (\bm{x},t)&= \frac{i\sqrt{\Sigma}}{2F}\int \frac{d^3k\sqrt{\bar{\omega}\left({\bm k}\right)}}{\left(2\pi\right)^3} \notag \\
&\quad \times \Big\{-\epsilon ^\alpha a_{II}(\bm{k})e^{-ik^{II}x}+\epsilon ^{*\alpha}a^\dagger_{II}(\bm{k})e^{ik^{II}x} 
\nonumber \\
&\quad -\epsilon ^{* \alpha} a_M(\bm{k})e^{-ik^Mx}+\epsilon ^{\alpha} a^\dagger_M(\bm{k})e^{ik^Mx} \Big\},\label{mom_in_feri} \\
\bar{\omega}\left({\bm k}\right)&\equiv\left(1+\left(\frac{2F^2}{v\Sigma}\right)^2\bm{k}^2\right)^{1/2}, \label{omegabar_in_feri}
\end{align}
where 
 $k^{II}x = \omega^{II}x_0 - {\bm k} \cdot {\bm x}$ and $k^{M}x=\omega^{M}x_0 - {\bm k} \cdot {\bm x}$ 
with $\omega^{II}$ and $\omega^{M}$ being defined
in Eq.~(\ref{dispersions}), and $(a^\dagger_{II},\,a_{II})$ and $(a^\dagger_{M},\,a_{M})$ 
are the corresponding creation and annihilation operators, satisfying the commutation relations,
\begin{align}
\left[a_{II}(\bm{k}),a_{II}^{\dagger}(\bm{q})\right]=&(2\pi)^3 \delta^3(\bm{k}-\bm{q}), \notag \\
\Big[a_{II}(\bm{k}),a_{II}(\bm{q})\Big]=&
\left[a_{M}^{\dagger}(\bm{k}),a_{II}^{\dagger}(\bm{q})\right]=0, \notag \\
\left[a_{M}(\bm{k}),a_{M}^{\dagger}(\bm{q})\right]=&(2\pi)^3 \delta^3(\bm{k}-\bm{q}), \notag \\
\Big[a_{M}(\bm{k}),a_{M}(\bm{q})\Big]=&
\left[a_{M}^{\dagger}(\bm{k}),a_{M}^{\dagger}(\bm{q})\right]=0, \notag \\
\Big[a_{II}(\bm{k}),a_{M}^\dagger(\bm{q})\Big]=&
\left[a_{M}(\bm{k}),a_{II}^{\dagger}(\bm{q})\right]=0. \label{com_IIandM}
\end{align}
 Note that two particle states may appear in contrast to the case of $1/v=0$. In this system, the vacuum is defined by
 \begin{align}
a_{II}(\bm{k})\left| 0 \right> =0, \notag \\
a_{M}(\bm{k})\left| 0 \right> =0, 
\end{align}
and the particle states are given by
\begin{align}
&\quad \left| \pi^{II}_{\bm{k}_1},\cdots,\pi^{II}_{\bm{k}_n},\pi^{M}_{\bm{k}_1},\cdots,\pi^{M}_{\bm{k}_m}\right> \nonumber \\
&\equiv a_{II}^\dagger(\bm{k}_1)\cdots  a_{II}^\dagger(\bm{k}_n)a_{M}^\dagger(\bm{k}_1)\cdots  a_{M}^\dagger(\bm{k}_m)\left| 0 \right>.
\end{align}

In the limit of vanishing $\Sigma$, Eqs.(\ref{NG_in_feri})-(\ref{omegabar_in_feri}) are reduced to
\begin{align}
\pi ^\alpha (\bm{x},t)&= \int \frac{d^3k}{\left(2\pi\right)^3}\sqrt{\frac{v}{2|{\bm k}|} }\notag \\
&\quad \times \Big\{\epsilon ^\alpha a_{II}(\bm{k})e^{-ik^{I}x}+\epsilon ^{*\alpha}a^\dagger_{II}(\bm{k})e^{ik^{I}x}  \nonumber \\
&\quad +\epsilon ^{* \alpha} a_M(\bm{k})e^{-ik^Ix}+\epsilon ^{\alpha} a^\dagger_M(\bm{k})e^{ik^Ix} \Big\},  \label{NG_in_antiferr} \\
P ^\alpha (\bm{x},t)&= \frac{i}{\sqrt{2}}\int \frac{d^3k}{\left(2\pi\right)^3}\sqrt{\frac{|{\bm k}|}{v}} \notag \\
&\quad \times \Big\{-\epsilon ^\alpha a_{II}(\bm{k})e^{-ik^{I}x}+\epsilon ^{*\alpha}a^\dagger_{II}(\bm{k})e^{ik^{I}x} 
\nonumber \\
&\quad -\epsilon ^{* \alpha} a_M(\bm{k})e^{-ik^Ix}+\epsilon ^{\alpha} a^\dagger_M(\bm{k})e^{ik^Ix} \Big\},\label{mom_in_antiferr} 
\end{align}
where $k^Ix=\omega ^I x_0 - {\bm k}\cdot {\bm x}$ with $\omega^I = v |{\bm k}|$. 
$a_{II}^\dagger$ and $a_M^\dagger$ create a particle with the same dispersion relation 
$\omega^I = v |{\bm k}|$, known as the type-I NG mode in the antiferromagnet. 
From now on, we change the notation of $a_{II}^\dagger$ and $a_M^\dagger$ ($a_{II}$ and $a_M$) to that of $a_1^\dagger$ and $a_2^\dagger$ ($a_1$ and $a_2$) for the magnons in the antiferromagnet.

\subsection{Charges, Order parameters, and Coefficients}
\label{condensates_and_coefficients}

To grasp the physical meaning of the coefficients in the effective Lagrangian, 
we discuss the relation among the charges, order parameters, 
and coefficients of the effective Lagrangian for the spin systems. 
The order parameter is defined as the expectation value of the commutation relation 
between the broken Noether charge $Q^\alpha$ and a local operator $\mathcal{O}(x)$,
\begin{align}
\lim _{V\rightarrow \infty}\frac{1}{V} \int _V d^3x \left<0\left| \left[ iQ^\alpha, \mathcal{O}(x) \right]\right| 0\right>,
\end{align}
where the commutation relation with the broken charge should be understood as
\begin{align}
\left[ iQ^\alpha, \mathcal{O}(x) \right]
\equiv \int d^{3}x^{\prime} \left[ ij_{0}^\alpha(x^{\prime}), \mathcal{O}(x) \right] .
\end{align}

The order parameter of the ferromagnet is the magnetization, 
which is given by the space integration of
the expectation value of the commutation relation 
between the Noether charge and the charge density, 
$\left<0\left| \left[ iQ^\alpha, j^\beta \right]\right| 0\right>$, 
which leads to the  quadratic dispersion relations of the NG modes 
and a reduction of the number of NG modes \cite{Leutwyler:1993gf, Hidaka:2012ym,Watanabe:2012hr}.
Furthermore, 
the coefficient $\Sigma$ 
in the effective Lagrangian of the ferromagnet is
identified with the magnetization \cite{Leutwyler:1993gf,Watanabe:2012hr},
\begin{align}
\Sigma = \lim _{V\rightarrow \infty}\frac{1}{V}\sum_m^N \left<0\right|  S^3_m \left| 0\right> ,\label{sigma_magnetization}
\end{align}
with $V$ being the total volume and $N$ the total number of lattice sites.

In this subsection, we first discuss that the relation is satisfied 
also in the case of the ferrimagnet, the effective Lagrangian for which 
contains terms with one and two time derivatives of the NG fields in
Eq.~(\ref{L_pi}). 
The charge densities $j_0 ^\alpha$ are derived from the effective Lagrangian,
\begin{align}
j_0^\alpha = \frac{\Sigma}{F}\epsilon^{\alpha\beta}\pi ^\beta + \frac{F}{v^2}\partial_0\pi^\alpha + O(\pi ^2).
\end{align}
We have used the transformation rule of the NG fields obtained from Eq.~(\ref{transformation_rule}),
\begin{align}
\delta ^\alpha \pi ^\beta =[iQ^{\alpha},\pi^{\beta}] = F\delta^{\alpha \beta} + O(\pi^\alpha) .
\end{align}
Note that the charge densities include the influence of the invariance only up to the total derivative in Eq.~(\ref{eff_L}). 
 Using the charge densities, the expectation value of the commutation relation 
between the charges and the charge densities reduces to
\begin{align}
&\quad \lim _{V\rightarrow \infty}\frac{1}{V}\int_V d^3x \left<0\left| \left[ iQ^\alpha, j_0^\beta \right]\right| 0\right> \nonumber \\
&=\lim _{V\rightarrow \infty}\frac{1}{V}\int_V d^3x \left<0\left|\left[iQ^\alpha,\frac{\Sigma}{F}\epsilon^{\beta\gamma}\pi ^\gamma + \frac{F}{v^2}\partial_0\pi^\beta\right]\right|0\right>\notag \\
&\simeq \lim _{V\rightarrow \infty}\frac{1}{V}\int_V d^3x\left<0\left|\frac{\Sigma}{F}\epsilon^{\beta \gamma}\delta ^{\alpha}\pi ^\gamma+ \frac{F}{v^2}\partial_0\left(\delta ^\alpha\pi^\beta\right)\right|0\right> \notag \\
&\simeq - \epsilon^{\alpha \beta}\Sigma .
\end{align}
This does not change by the redefinition of the NG fields,
$\pi^\alpha \rightarrow \pi^\alpha/Z$, with a constant $Z$. 
By matching the charge density in an underlying theory, the expectation value of the commutation relation is also rewritten as
\begin{align}
&\quad \lim _{V\rightarrow \infty}\frac{1}{V}\int_V d^3x \left<0\left| \left[ iQ^\alpha, j_0^\beta \right]\right| 0\right> \nonumber \\
&
=-\epsilon^{\alpha \beta}\lim _{V\rightarrow \infty}\frac{1}{V}\int_V d^3x \left<0\left|  j_0^3 \right| 0\right>\notag \\
&=-\epsilon^{\alpha \beta}\lim _{V\rightarrow \infty}\frac{1}{V}\sum_m ^N\left<0\left|  S_m^3 \right| 0\right>.
\end{align}
Thus, the coefficient $\Sigma$ is identified with the magnetization also in the case of the ferrimagnet; see Eq.~(\ref{sigma_magnetization}). 
Note that in the case of the antiferromagnet, the magnetization vanishes, $\Sigma=0$, 
and the effective Lagrangian does not involve terms with one time derivative\cite{Leutwyler:1993gf, Hofmann:1998pp}.

We may consider the commutation relation between the Noether charge and the NG fields:
\begin{align}
&\quad \lim _{V\rightarrow \infty}\frac{1}{V}\int _V d^3x \left<0\left| \left[ iQ^\alpha, \pi^\beta \right]\right| 0\right> \nonumber \\
&=\lim _{V\rightarrow \infty}\frac{1}{V}\int_V d^3x \left<0\left|\delta^\alpha \pi^\beta \right|0\right>\notag \\
&=\delta^{\alpha \beta}F. \label{Noether_and_NG}
\end{align}
The relation changes by redefinition of the NG fields. 
If we redefine the NG fields as $\pi^\alpha \rightarrow \left(F/F_t\right)\pi^\alpha $, 
the relation changes and is given in terms of the coefficient of the terms with two time derivatives of the
NG fields in the effective Lagrangian as 
\begin{align}
\lim _{V\rightarrow \infty}\frac{1}{V}\int_V d^3x \left<0\left| \left[ iQ^\alpha, \pi^\beta \right]\right| 0\right>=\delta^{\alpha \beta}F_t.
\end{align}
 Because $1/v=F_{t}/F$, the nonzero expectation value leads to another order parameter in the case 
of the EFT containing terms with one and two time derivatives of the NG fields, 
which leads to the appearance of staggered magnetization in the ferrimagnet. 
The relation Eq.~(\ref{Noether_and_NG}) is also satisfied 
in the case of the EFT with one time derivative but without two time derivatives, 
where, however, 
 we cannot distinguish the NG field $\pi ^\alpha$ from the charge density $j_0 ^\alpha$. 
In fact, an explicit calculation using Eq.~(\ref{NG_in_fero})
shows that $\int d^3x \pi^\alpha $ is conserved; 
\begin{align}
&\quad \frac{d}{d t}\int d^3x \pi^\alpha \nonumber \\
&=\int d^3x
\frac{F}{\sqrt{\Sigma}}\int \frac{d^3k}{\left(2\pi\right)^{3}}\nonumber \\
&\quad \times\left\{\epsilon^\alpha\left(-i\omega\right) a(\bm{k}) e^{-ikx}+\epsilon ^{\ast \alpha} \left(i\omega\right)a^{\dagger}(\bm{k}) e^{ikx}\right\} \notag \\
&=\frac{F}{\sqrt{\Sigma}}\int d^3k\delta\left({\bm k}\right)\nonumber \\
&\quad \times\left\{\epsilon^\alpha\left(-i\omega\right) a(\bm{k}) e^{-i\omega t}+\epsilon ^{\ast \alpha} \left(i\omega\right)a^{\dagger}(\bm{k}) e^{i\omega t}\right\} \notag \\
&=0,
\end{align}
which does not vanish in the case of the EFT with 
both one and two time derivatives of the NG fields, i.e., 
$\frac{d}{d t}\int d^3x \pi ^\alpha \neq 0$. 
Thus in the case of a ferromagnet, we have only one order parameter given by the magnetization. 
In other words, the effective Lagrangian for the ferromagnet 
does not have terms with two time derivatives even 
if we take into account the contribution from higher derivatives. 
Similar relations have been obtained for a general system on the basis  
of the Langevin equation \cite{Hayata:2014yga}.  

\subsection{Scattering amplitudes and the scattering lengths}
In this subsection, we calculate the scattering amplitudes between two NG modes 
at tree level, and their scattering lengths. 
These are derived from the $\pi^{4}$ terms in Eq.~\eqref{L_pi}. 
Corresponding to the case of (a) ferromagnet ($1/v=0, \Sigma \neq 0$), 
(b) antiferromagnet ($1/v\ne 0, \Sigma = 0$), and (c) ferrimagnet ($1/v\ne 0, \Sigma \neq 0$), 
we discuss the amplitudes and the scattering lengths, respectively. 
As for the cases of ferromagnet and antiferromagnet, a similar 
calculation
was already performed 
using different coordinates system describing the EFT \cite{Hofmann:1998pp}: 
In fact, the form of the resulting Lagrangian is different from ours given in Eqs.~(\ref{L_pi})-(\ref{L02}), 
and our calculation ensures that the scattering amplitudes do not depend on the 
choice of the coordinates. 
Furthermore, our calculation includes the first study on the case of the ferrimagnet.

(a) In the case of the ferromagnet where $1/v=0, \Sigma \neq 0$, 
only one magnon mode with the dispersion relation $\omega = \frac{F^2}{\Sigma}{\bm k}^2$ appears. 
We denote the particle with momentum ${\bm k}$ by $\pi ({\bm k})$. 
A scattering process for $\pi({\bm k_1})+\pi({\bm k_2}) \rightarrow \pi({\bm k_3} )+\pi({\bm k_4})$ occurs, and the scattering amplitude is given by
\begin{align}
&M\left[\pi({\bm k_1})+\pi({\bm k_2}) \rightarrow \pi({\bm k_3} )+\pi({\bm k_4})\right] \notag \\
&=-\frac{1}{6\Sigma}\left(\omega_1+\omega_2+\omega_3+\omega_4\right)\nonumber \\
&\quad +\frac{F^2}{3\Sigma^2}\bigl[{\bm k}_1 \cdot {\bm k}_3+{\bm k}_2 \cdot {\bm k}_3+{\bm k}_1 \cdot {\bm k}_4+{\bm k}_2 \cdot {\bm k}_4 \nonumber \\
&\quad + 2\left({\bm k}_1 \cdot {\bm k}_2+{\bm k}_3 \cdot {\bm k}_4\right)\bigr] \notag \\
&=\frac{2F^2}{\Sigma^2}\left({\bm k}_1 \cdot {\bm k}_2 \right),
\end{align}
where we have used momentum and energy conservation, 
${\bm k}_1+ {\bm k}_2={\bm k}_3 + {\bm k}_4$, and $\omega_1 + \omega_2=\omega_3+\omega_4$ with $\omega_i = \frac{F^2}{\Sigma}{\bm k}_i^2$.
This result coincides with the previous study \cite{Hofmann:1998pp}, 
and is in accordance with Dyson's microscopic analysis \cite{Dyson:1956zz}. 

The definition of the scattering length and phase shift through the scattering amplitude in the nonrelativistic notation are given 
in Appendix \ref{scattering length}. In the case of the ferromagnet system without external fields, the scattering length vanishes, though it may not be well-defined in the process, because of the appearance of scattering processes with massless particles. This is a consequence of the derivative couplings of EFT, in accordance with the Goldstone theorem.

(b) In the case of the antiferromagnet where $1/v\neq 0, \Sigma=0$, 
we have two magnons with the same dispersion relation $\omega ^I= v|{\bm k}|$ 
as mentioned in \ref{Construction of the Hamiltonian and particle states}. 
Let $\pi^ \alpha ({\bm k}) (\alpha=1,2)$ denote the particle states 
created by $a_\alpha ^\dagger$. 
The unbroken SO(2) symmetry leads to the conservation of the number of particles, 
and only the following three scattering processes are allowed: 
$\pi^1({\bm k}_1)+\pi^1({\bm k}_2) \rightarrow \pi^1({\bm k}_3)+\pi^1({\bm k}_4), \pi^1({\bm k}_1)+\pi^2({\bm k}_2) \rightarrow \pi^1({\bm k}_3)+\pi^2({\bm k}_4),$ and $\pi^2({\bm k}_1)+\pi^2({\bm k}_2) \rightarrow \pi^2({\bm k}_3)+\pi^2({\bm k}_4)$. The scattering amplitudes are given by
\begin{align}
&M\left[\pi ^{1}({\bm k}_1)+\pi ^{1}({\bm k}_2) \rightarrow \pi ^{1}({\bm k}_3)+\pi ^{1}({\bm k}_4)\right] \notag \\
&=M\left[\pi ^{2}({\bm k}_1)+\pi ^{2}({\bm k}_2) \rightarrow \pi ^{2}({\bm k}_3)+\pi ^{2}({\bm k}_4)\right] \notag \\
&= -\frac{v^2}{2F^2\sqrt{|{\bm k}_1||{\bm k}_2||{\bm k}_3||{\bm k}_4|}} 
\left(|{\bm k}_1||{\bm k}_2|-{\bm k}_1\cdot{\bm k}_2\right) \\
&M\left[\pi ^{1}({\bm k}_1)+\pi ^{2}({\bm k}_2) \rightarrow \pi ^{1}({\bm k}_3)+\pi ^{2}({\bm k}_4)\right] \notag \\
&= \frac{v^2}{2F^2\sqrt{|{\bm k}_1||{\bm k}_2||{\bm k}_3||{\bm k}_4|}} 
\left(|{\bm k}_2||{\bm k}_3|-{\bm k}_2\cdot{\bm k}_3\right),
\end{align}
where we have used momentum and energy conservation. \footnote{The scattering amplitudes in Ref.\cite{Hofmann:1998pp} are obtained by changing the definition of the particle states. The creation operator in Ref.\cite{Hofmann:1998pp} corresponds to $\tilde{a}_1^{\dagger} =\left(a_1^\dagger+a_2^\dagger \right)/\sqrt{2}$ and $\tilde{a}_2^\dagger =i\left(a_1^\dagger-a_2^\dagger \right)/\sqrt{2}$.} Also in the case of the antiferromagnet, the scattering length vanishes. 

(c) In the case of the ferrimagnet where $1/v\neq0, \Sigma \neq 0$, 
 we also have two magnons with the dispersion relations Eq.(\ref{dispersions}), 
denoted by $\pi^{II}({\bm k})$ and $\pi^{M}({\bm k})$. 
The following three scattering processes are allowed; 
$\pi ^{II}({\bm k_1})+\pi ^{II}({\bm k_2}) \rightarrow \pi ^{II}({\bm k_3})+\pi ^{II}({\bm k_4}), \pi ^{II}({\bm k_1})+\pi ^{M}({\bm k_2}) \rightarrow \pi ^{II}({\bm k_3})+\pi ^{M}({\bm k_4}) ,$ and $\pi ^{M}({\bm k_1})+\pi^{M}({\bm k_2}) \rightarrow \pi ^{M}({\bm k_3})+\pi ^{M}({\bm k_4})$. Other processes such as
$\pi^{II}({\bm k_1})+\pi^{II}({\bm k_2}) \rightarrow \pi^M({\bm k_3})+\pi^M({\bm k_4})$ 
are prohibited because of the residual symmetry SO(2), as is the case for the antiferromagnet.

The scattering amplitudes are as follows:
\begin{align}
&M\left[\pi ^{II}({\bm k_1})+\pi ^{II}({\bm k_2}) \rightarrow \pi ^{II}({\bm k_3})+\pi ^{II}({\bm k_4})\right] \notag \\
&=\frac{1}{\sqrt{\bar{\omega } _1\bar{\omega}  _2 \bar{\omega}  _3 \bar{\omega } _4}}\Bigl[-\frac{1}{3\Sigma}\left(\omega_1^{ II}+\omega_2^{ II}\right) \nonumber \\
&\quad +\frac{F^2}{3\Sigma ^2}\left(4\bm{k}_1\cdot\bm{k}_2 +2\bm{k}_3\cdot\bm{k}_4 +\bm{k}_1^2+\bm{k}_2^2 \right)\Bigr], \\
&M\left[\pi ^{II}({\bm k_1})+\pi ^{M}({\bm k_2}) \rightarrow \pi ^{II}({\bm k_3})+\pi ^{M}({\bm k_4})\right] \notag \\
&=\frac{1}{\sqrt{\bar{\omega } _1\bar{\omega}  _2 \bar{\omega}  _3 \bar{\omega } _4}}\Bigl[\frac{1}{6\Sigma}\left(\omega_2^{ M}+\omega_4^{ M}-\omega_1^{ II}-\omega_3^{ II}\right)\nonumber \\
&\quad -\frac{F^2}{3\Sigma ^2}\left(4\bm{k}_2\cdot\bm{k}_3 +2\bm{k}_1\cdot\bm{k}_4- \bm{k}_2^2-\bm{k}_3^2  \right)\Bigr], \\
&M\left[\pi ^{M}({\bm k_1})+\pi^{M}({\bm k_2}) \rightarrow \pi ^{M}({\bm k_3})+\pi ^{M}({\bm k_4})\right] \notag \\
&=\frac{1}{\sqrt{\bar{\omega } _1\bar{\omega}  _2 \bar{\omega}  _3 \bar{\omega } _4}}\Bigl[\frac{1}{3\Sigma}\left(\omega_1^{ M}+\omega_2^{ M}\right)\nonumber \\
&\quad +\frac{F^2}{3\Sigma ^2}\left(4\bm{k}_1\cdot\bm{k}_2 +2\bm{k}_3\cdot\bm{k}_4 +\bm{k}_1^2+\bm{k}_2^2  \right)\Bigr],
\end{align}
with $\bar{\omega}_i \equiv \left[1+\left(\frac{2F^2}{v\Sigma}\right)^2\bm{k}_i^2\right]^{1/2},
\omega^{II}_i\equiv \frac{v^2\Sigma}{2F^2}\left(-1+\bar{\omega}_i\right)$, and
$\omega^{M}_i\equiv \frac{v^2\Sigma}{2F^2}\left(1+\bar{\omega}_i \right)~(i=1,\dots ,4)$ .

The scattering lengths for $\pi^{II}({\bm k_1})+\pi^{M}({\bm k_2}) \rightarrow \pi^{II}({\bm k_3})+\pi^M({\bm k_4})$ and $\pi^{M}({\bm k_1})+\pi^{M}({\bm k_2}) \rightarrow \pi^{M}({\bm k_3})+\pi^M({\bm k_4})$ are finite in contrast to the ferromagnet case:
\begin{align}
&a^{II+M\rightarrow II+M}=\frac{\Sigma}{12F^2 F_t^2}=\frac{\nu_{M}}{12F^2}, \label{scatteringlengthIIM} \\
&a^{M+M\rightarrow M+M}=\frac{\Sigma}{6F^2 F_t^2}=\frac{\nu_{M}}{6F^2 },
\label{scatteringlengthMM}
\end{align}
where the mass gap is defined as $\nu_{M}=\Sigma/F_{t}^{2}$. 
Because $\Sigma$ is identified with the magnetization, 
$F^{2}$ can be determined experimentally 
by measuring the $\bm{k}^{2}$ dependence of the dispersion relation of the gapped NG mode 
through Eq.~\eqref{dispersions}. 
In this way, the magnitude of the scattering length is used
to determine 
the coefficients in the dispersion relation. 
Moreover, the positive sign of the scattering lengths $a^{II+M\rightarrow II+M},a^{M+M\rightarrow M+M}>0$ 
meaning an attractive interaction 
and the ratio $a^{II+M\rightarrow II+M}/a^{M+M\rightarrow M+M}=1/2$ are independent of the low energy constants. 
These properties are consequences of the low energy theorem for the gapped NG mode.

\section{Analysis in the presence of external fields: magnetic fields and anisotropic effects}
\label{EFT with external fields}

In chiral perturbation theory, 
 terms involving explicit symmetry breaking can be introduced 
by assuming appropriate transformation properties 
of the symmetry-breaking terms so as to make the effective Lagrangian invariant under chiral symmetry.
In much the same way, we can  involve
 the explicit breaking in the EFT for magnons. 
First we introduce the explicit symmetry-breaking terms in the underlying theory, 
in terms of
 an external magnetic field and single-ion anisotropy:
\begin{align}
\mathcal{H}=\mathcal{H}_{\mathrm{inv}}(S_{m}^{a}) -\mu H^a \sum_m S_m^a-D^{ab}\sum_m  S^a_mS^b_m,
\label{Hexternal}
\end{align}
where $\mathcal{H}_{\mathrm{inv}}$ is the invariant term under a rotation 
of the spins, and the terms including $H$ and $D^{ab}$  
represent the effect of 
 the magnetic field and single-ion anisotropy, respectively. 
When $H=D^{ab}=0$, as discussed in Sec.~\ref{EFT}, 
the underlying theory is invariant under SO(3), and 
the symmetry is broken down to SO(2) in the ground state. 
When the magnetic field is applied, the spins couple to the magnetic field, 
and the explicit symmetry breaking occurs. 
We assume that the direction of the magnetic field is in the 3-axis, $H^a = \delta^{a3}H$. 
Also the anisotropic effect leads to an explicit breaking, 
and the squared spins on the same site, $S^a_mS^b_m$ couple to the matrix $D^{ab}$. 
Now we consider the diagonal case $D^{ab}= D^a \delta^{ab}$ for simplicity.

Let us first discuss the effect 
of the magnetic field on the EFT;see also
\cite{Leutwyler:1993gf, Hofmann:1998pp, roman9709298effective} where the cases 
are discussed. 
In Eq.~\eqref{Hexternal}, the magnetic field couples to the SO(3) charges, 
$\sum_m S_m^a\sim \int d^3x j_0^a(x)$, which leads to the identification 
of the magnetic fields with the time component of SO(3) gauge fields 
$A_0 \left(x\right)=A_0^a\left(x\right)T^a$. In fact, if we consider 
the continuum limit, the term is reduced
to $\int d^3x A_0^a(x) j_0^a (x)$, which is invariant for the partial gauge transformation for SO(3),
\begin{align}
A_0 \left( x\right) \rightarrow g \left(t\right)A_0 \left(x\right)g^{-1} \left(t\right)+ 
\frac{1}{i} g \left(t\right) \partial _0 g^{-1} \left(t\right) ,
\end{align}
with $g \left(t\right) =e^{i\theta^3\left(t\right) T^3} \in$ SO(2). 
Thus, the magnetic fields are handled 
in the EFT by the identification $A_0^a (x) = \delta^{a 3}H$ 
after we construct the SO(3) gauge-invariant effective Lagrangian only up 
to a total derivative \cite{Leutwyler:1993gf}. 
In practice, this is easily performed by replacing 
the time derivatives with the following form in Eq.~(\ref{eff_L}),
\begin{align}
\partial_0 \rightarrow \partial_0-i \mu H T^3.
\label{magnetic}
\end{align}

Next, we discuss the effect of the single-ion anisotropy. 
In the Heisenberg model, the transformation of the coefficient $D^{ab}$ is considered as
\begin{align}
D^{ab} \rightarrow \left(g D g^{-1}\right)^{ab} ,
\end{align}
with the global transformation $g\in$ SO(3). 
In the EFT, there are the following invariant terms at leading order $O(D^{ab})$:
\begin{align}
&\alpha _1 [U^{-1}DU]^{33} \simeq \alpha_1 D_{33}+ \cdots, \\ 
&\alpha_2 [U^{-1}DU]^{\alpha \alpha} \simeq \alpha _2  D_{\alpha \alpha}+ \cdots, \\ 
&\alpha _3 \mathrm{Tr}\left[T^3U^{-1} D U \right]\simeq \alpha _3i\epsilon _{\alpha \beta}  D_{\alpha \beta}+ \cdots ,
\end{align}
where we have shown the invariance of the first term by using 
$h^{3a}=\left[e^{i\theta T^3}\right]^{3a}=\delta^{3a}$. Note that the last term, 
$\mathrm{Tr}\left[T^3U^{-1} D U \right]$, is always zero in the present case with 
$D_{ab}=D_{a}\delta_{ab}$. The coefficients $\alpha _1$ and $\alpha _2$ are 
determined by matching the condensates in the underlying theory without external fields:
\begin{align}
\lim_{V\rightarrow \infty} \frac{1}{V}\sum_m^N \left< S^3_m S^3_m \right> 
&=\left. \lim_{V\rightarrow \infty} \frac{1}{V}\frac{\delta Z_{\rm eff}(H,D)}{\delta D^{33}}\right|_{H=0,D=0}, \\
\lim_{V\rightarrow \infty} \frac{1}{V}\sum_m^N \left< S^\alpha_m S^\beta_m \right> 
&=\left. \lim_{V\rightarrow \infty} \frac{1}{V}\frac{\delta Z_{\rm eff}(H,D)}{\delta D^{\alpha\beta}}\right|_{H=0,D=0},
\end{align}
with $Z_{\rm eff}(H,D)$ being the partition function of the effective field theory 
with external fields, $N$ the total number of lattice sites, and $V$ the total volume. 

We give several examples of the underlying theories as a summary. 
The simplest case is a spin-1/2 quantum Heisenberg model~\eqref{eq:Heisenberg}, in which the condensates are given by
\begin{align}
&\sum_m \left< S^3_m S^3_m \right>=\frac{N}{4},  \\
&\sum_m \left< S^\alpha_m S^\beta_m \right>= \delta ^{\alpha\beta}\frac{N}{4}+\frac{i\epsilon ^{\alpha\beta}}{2}\sum _{m }\left< S_m^3 \right>,
\end{align}
which leads to
\begin{align}
\alpha_1=\alpha_2\equiv \alpha.
\end{align}
The term involving the anisotropic effect is reduced to
\begin{align}
\mathcal{L}^D&=\alpha \mathrm{Tr}U^{-1}DU =\alpha \mathrm{Tr}D.
\end{align}
Thus in the case of the spin-1/2 Heisenberg model, 
the anisotropic effects do not appear in the low-energy regime. 
Note that the result is 
readily obtained in the conventional Heisenberg model because 
$\sum_a D^a S^a S^a=\sum_a D^a/4$. 
What we have shown is that
the result holds in a model-independent way 
for any spin-1/2 underlying theories 
 having the same expectation values of the spin operators.

In a spin-S Heisenberg ferromagnet, the condensates are given by
\begin{align}
&\sum_m \left< S^3_m S^3_m \right> 
= S^2 N, \\
&\sum_m \left< S^\alpha_m S^\beta_m \right>
=\frac{\delta^{\alpha\beta}}{2}SN+\frac{i\epsilon ^{\alpha\beta}}{2}SN,
\end{align}
and the coefficients $\alpha_1$ and $\alpha _2$ are found to be
\begin{align}
\alpha _1 = \frac{N}{V}S^2, \quad \alpha _2 &= \frac{N}{2V}S.
\end{align}
Note that the inequality $\alpha _1 > \alpha _2$ 
except for the case of a spin-1/2 ferromagnet, 
and the coefficients are related to the magnetization as $\Sigma =\frac{N}{V}S$.

In a spin-$(S_A, S_B)$ Heisenberg ferrimagnet, the condensates are given by
\begin{align}
&\sum_m \left< S^3_m S^3_m \right> 
= \frac{N}{2}(S_A^2 + S_B^2), \\
&\sum_m \left< S^\alpha_m S^\beta_m \right>
=\frac{\delta^{\alpha\beta}}{4}N(S_A + S_B)+\frac{i\epsilon^{\alpha\beta}}{4}N(S_A - S_B),
\end{align}
and the coefficients $\alpha_1$ and $\alpha _2$ are found to be 
\begin{align}
\alpha _1 &= \frac{N}{2V}(S_A^2 + S_B^2).  \\
\alpha _2 &= \frac{N}{4V}(S_A + S_B). 
\end{align}
As in the ferromagnetic case, the inequality $\alpha _1 > \alpha _2$ holds
except for the case 
with $S_{A}=S_{B}=1/2$, i.e., a spin-1/2 antiferromagnet. 
Using the explicit expressions of the magnetization $\Sigma =(S_A-S_B)N/2V$ and the staggered magnetization $\Sigma _h=(S_A+S_B)N/2V,$ the coefficients are rewritten as $\alpha _1=(\Sigma ^2+ \Sigma _h^2)V/N$ and $\alpha _2=\Sigma _h /2$.

In terms of the NG fields $\pi ^\alpha$, 
the effective Lagrangian involving the magnetic field and anisotropic effect is expressed as
\begin{align}
\mathcal{L} &= \mathcal{L}_{2\pi}+\mathcal{L}_{4\pi}+\dotsb \notag \\
\mathcal{L}_{2\pi} &= -\frac{\Sigma}{2F^2}\epsilon ^{\alpha \beta}\pi^\alpha D_0 \pi ^\beta +\frac{1}{2v^2} D_0 \pi ^\alpha D_0 \pi ^\alpha -\frac{1}{2}\partial_i \pi ^\alpha \partial_i \pi ^\alpha \notag \\
&\quad +\frac{D_2-D_3}{F^2}\left(\alpha _1-\alpha _2 \right)\pi^1\pi^1 +\frac{D_1-D_3}{F^2}\left(\alpha _1-\alpha _2 \right)\pi^2\pi^2,\notag \\
\mathcal{L}_{4\pi}&= \frac{\Sigma}{24F^4}\epsilon^{\alpha \beta}\pi ^\alpha D_0 \pi ^\beta \pi^\gamma \pi^\gamma \nonumber\\
&\quad -\frac{1}{6v^2F^2}\left[D_0 \pi^\alpha D_0 \pi^\alpha \pi^\beta \pi^\beta-\pi^\alpha D_0\pi^\alpha \pi^\beta D_0 \pi^\beta\right]\notag \\
&\quad+\frac{1}{6F^2}\left[\partial_i \pi^\alpha \partial_i \pi^\alpha \pi^\beta \pi^\beta-\pi^\alpha\partial_i\pi^\alpha \pi^\beta \partial_i \pi^\beta\right] \notag \\
&\quad+ \frac{D_3-D_1}{3F^4}\left(\alpha _1-\alpha _2\right)\pi ^\alpha\pi^\alpha \pi^2\pi^2 \nonumber \\
&\quad + \frac{D_3-D_2}{3F^4}\left(\alpha _1-\alpha _2\right)\pi ^\alpha\pi^\alpha \pi^1\pi^1 ,
\label{L4pi}
\end{align}
with the covariant derivative for the time component $D_0 \pi ^\alpha =\left(\partial_0 \delta^{\alpha \beta}-\epsilon ^{\alpha \beta}\mu H \right) \pi ^\beta$. We neglect the terms only with external fields, $H$ and $D$. Because we assume that the spins are (anti-)aligned in the 3-axis direction, the anisotropy effect should be maximum in the 3-direction, $D_{3}\gg D_{1},D_{2}$. In this section, we show the magnon-magnon scattering in the presence of the only 3-axis anisotropic effect $D_1=D_2=0, D_3 \neq 0$. As in Sec.~\ref{EFT}, we classify the cases by the value of $1/v$.

We now consider the power counting scheme with the external fields.
 For a reasonable application of EFT, a definite power counting scheme should be established 
for all the fields including the external fields, for which 
the strength of the external fields should 
be constrained. 
Because the external magnetic field is introduced in the form of Eq.~\eqref{magnetic}, 
its strength should be $\mu H\sim \partial_{0}\ll \nu$ where $\nu$ 
is the first gapped excitation other than the NG modes. 
When $\Sigma\neq 0$, because the anisotropy effect provides the gap 
$(\alpha_{1}-\alpha_{2})D_{3}/\Sigma$, this quantity should also be of the order of $\partial_{0}$. 
We thus consider that the energy scale of the external field $E$ is counted as the same order 
as that of the NG boson field as
\begin{align}
&E=\mu H , \frac{\alpha_{1}-\alpha_{2}}{\Sigma}D_{3} \\
&
\begin{cases}
\frac{\partial_{0}}{\nu}\sim\frac{E}{\nu} \sim (\frac{\partial_{i}}{\Lambda})^{2} 
\sim p^{2} & 1/v= 0, \Sigma\neq 0 \\
\frac{\partial_{0}}{\nu}\sim\frac{E}{\nu}\sim \frac{\nu_{M}}{\nu} \sim  (\frac{\partial_{i}}{\Lambda})^{2}
\sim p^{2} & 1/v\neq  0, \Sigma\neq 0 
\end{cases} .
\label{counting}
\end{align}
Analyzing the dispersion relation for $\Sigma=0$ in the same manner, we obtain the counting rule as \begin{align}
(\tfrac{\partial_{0}}{\nu})^{2} 
\sim(\tfrac{\mu H}{\nu})^{2}
\sim(\tfrac{\partial_{i}}{\Lambda})^{2} 
\sim \tfrac{(\alpha_{1}-\alpha_{2})D_{3}}{F^{2}}
\sim p^{2} &\quad  1/v\neq 0, \Sigma=0  .
\label{counting2}
\end{align}

(a) In the case of $1/v = 0$ and $\Sigma \neq 0$, the asymptotic fields involving the external fields are given by
\begin{align}
\pi ^\alpha(\bm{x},t)&=\frac{F}{\sqrt{\Sigma}}\int \frac{d^3k}{\left(2\pi\right)^{3}}\left\{\epsilon^\alpha a(\bm{k}) e^{-ikx}+\epsilon ^{\ast \alpha} a^{\dagger}(\bm{k}) e^{ikx}\right\}, \notag \\
\omega&\equiv 
\mu H
+\frac{\alpha_1 -\alpha _2}{\Sigma}D_3
+\frac{F^2}{\Sigma}\bm{k}^2. 
\end{align}
We note that the NG mode acquires the mass gap which is proportional to the external fields, $H$ and $D_{3}$. The amplitude for the scattering process, $\pi ({\bm k_1})+\pi ({\bm k_2}) \rightarrow \pi ({\bm k_3})+\pi ({\bm k_4})$, is given by
\begin{align}
&M\left[\pi({\bm k_1})+\pi({\bm k_2}) \rightarrow \pi({\bm k_3})+\pi({\bm k_4})\right] 
\nonumber \\
&=\frac{2F^2}{\Sigma^2}\left({\bm k}_1 \cdot {\bm k}_2 \right)+\frac{14}{3}\left(\alpha_1-\alpha_2\right)\frac{D_3}{\Sigma^2}
\end{align}
Note that the scattering amplitude does not depend on the magnetic field $H$, but
on the anisotropic effect $D_{3}$. The scattering length, which is defined in Appendix \ref{scattering length} 
reads:
\begin{align}
a = \frac{7}{6}\left(\alpha_1-\alpha_2\right)\frac{D_3}{\Sigma F^2},
\end{align}
which is finite and proportional to the anisotropy $D_{3}$, as long as $\alpha_{1}\neq \alpha_{2}$. 
In QCD, pion scattering length is proportional to the pion mass squared \cite{Weinberg:1966kf}, implying that the nonzero scattering length is caused by
the explicit symmetry breaking due to the current quark masses in QCD.
In the magnon scattering, we consider two types of the explicit symmetry breaking, the external magnetic
 field $H$ and the anisotropy $D_{3}$. While both effects contribute to the mass gap in Eq.~(95),
 the scattering length depends only on the anisotropy $D_{3}$. This is because of the difference
 in the breaking patterns of the SO(3) symmetry. In the spin system, it is the anisotropy $D_{3}$
 that corresponds to the explicit symmetry breaking by the quark mass term in QCD.

If we write the mass gap by the anisotropic effect 
as $\nu_{D}=(\alpha_{1}-\alpha_{2})D_{3}/\Sigma$, we obtain
\begin{align}
a = \frac{7\nu_{D}}{ 6F^2} .
\end{align}
This is an analogous relation to Eqs.~\eqref{scatteringlengthIIM} and \eqref{scatteringlengthMM}. 
Note, however, that the coefficient $7/6$ is an order of magnitude larger than those 
of the gapped NG modes, $1/12$ and $1/24$. If the anisotropic effect induces a comparable amount 
of the gap $\nu_{D}$ with the gapped modes, the scattering length should be much larger.

(b) In the case of $1/v \neq 0$ and $\Sigma=0$, when the external fields are applied, the asymptotic NG fields are expanded as follows:
\begin{align}
\pi ^\alpha (\bm{x},t)&= \int \frac{d^3k}{\left(2\pi\right)^3}\sqrt{\frac{v}{2\bar{k}} }\notag \\
&\quad \times \Big\{\epsilon ^\alpha a_{1}(\bm{k})e^{-ik^{1}x}+\epsilon ^{*\alpha}a^\dagger_{1}(\bm{k})e^{ik^{1}x}  \nonumber \\
&\quad +\epsilon ^{* \alpha} a_2(\bm{k})e^{-ik^2x}+\epsilon ^{\alpha} a^\dagger_2(\bm{k})e^{ik^2x} \Big\}, \\
&\bar{k} \equiv \sqrt{{\bm k}^2+\frac{D_3}{F^2}\left(\alpha_1 -\alpha_2\right)} \label{NG_in_antiferr_with_extenal} ,
\end{align}
where 
$k^\alpha x=\omega^\alpha x_0-{\bm k}\cdot {\bm x}$ with
\begin{align}
\omega^1 &=\mu H + v\bar{k}, \notag \\
\omega^2 &=-\mu H + v\bar{k}. \label{dispersion_antiferro_external}
\end{align}

Note that $\bar{k}$ does not depend on the magnetic field. 
The $\pi ^\alpha ({\bm k})$ denote 
the particle states with the dispersion relation $\omega ^\alpha$ created by $a^\dagger _\alpha$. 
The scattering amplitudes for $\pi^1 ({\bm k}_1)+ \pi^1  ({\bm k}_2)\rightarrow \pi^1  ({\bm k}_3)+ \pi^1 ({\bm k}_4), \pi^2 ({\bm k}_1) + \pi^2 ({\bm k}_2) \rightarrow \pi^2 ({\bm k}_3) + \pi^2 ({\bm k}_4),$ and $\pi^1  ({\bm k}_1)+ \pi^2  ({\bm k}_2)\rightarrow \pi^1 ({\bm k}_3) + \pi^2 ({\bm k}_4)$ are given by 
\begin{align}
&M\left[\pi ^{1}({\bm k}_1)+\pi ^{1}({\bm k}_2) \rightarrow \pi ^{1}({\bm k}_3)+\pi ^{1}({\bm k}_4)\right] \notag \\
&=M\left[\pi ^{2}({\bm k}_1)+\pi ^{2}({\bm k}_2) \rightarrow \pi ^{2}({\bm k}_3)+\pi ^{2}({\bm k}_4)\right] \notag \\
&= -\frac{v^2}{6F^2\sqrt{\bar{k}_1\bar{k}_2\bar{k}_3\bar{k}_4}} 
\Bigl[3\left(\bar{k}_1\bar{k}_2-{\bm k}_1\cdot{\bm k}_2\right) \notag \\
&-7\left(\alpha_1-\alpha_2\right)\frac{D_3}{F^2}\Bigr] \\
&M\left[\pi ^{1}({\bm k}_1)+\pi ^{2}({\bm k}_2) \rightarrow \pi ^{1}({\bm k}_3)+\pi ^{2}({\bm k}_4)\right] \notag \\
&= \frac{v^2}{6F^2\sqrt{\bar{k}_1\bar{k}_2\bar{k}_3\bar{k}_4}} 
\Bigl[3\left(\bar{k}_1\bar{k}_3-{\bm k}_1\cdot{\bm k}_3\right) \notag \\
&+7\left(\alpha_1-\alpha_2\right)\frac{D_3}{F^2}\Bigr] .
\end{align}
Note that the scattering amplitudes do not depend on the magnetic field, 
as in the case of the ferromagnet. A different result in a previous study \cite{Hofmann:1998pp} is due to the dependence of the magnetic field for $\bar{k}$ in Eq.(\ref{NG_in_antiferr_with_extenal}). 
Our result for $\bar{k}$ is consistent with the quantization conditions (\ref{com_pp}) - (\ref{com_PP}). The scattering lengths are given by
\begin{align}
a^{1+1 \rightarrow  1+1}&=a^{2+2 \rightarrow 2+2} =\frac{\nu _D}{3F^2} \\
a^{1+2 \rightarrow  1+2}&=\frac{5\nu _D}{6F^2}
\end{align}
with the mass gap $\nu_D =v\sqrt{(\alpha_1-\alpha_2)D_3/F^2}.$ 

(c) In the case of $1/v \neq 0$ and $\Sigma \neq 0$, the asymptotic fields involving the external fields are given by
\begin{align}
\pi ^\alpha (\bm{x},t)&= \frac{F}{\sqrt{\Sigma}}\int \frac{d^3k} {\left(2\pi\right)^3\sqrt{\bar{\omega}\left({\bm k}\right)}}
\nonumber \\
&\quad \times 
\Big\{\epsilon ^\alpha a_{II}(\bm{k})e^{-ik^{II}x}+\epsilon ^{*\alpha}a^\dagger_{II}(\bm{k})e^{ik^{II}x}  \notag \\
&\quad +\epsilon ^{* \alpha} a_M(\bm{k})e^{-ik^Mx}+\epsilon ^{\alpha} a^\dagger_M(\bm{k})e^{ik^Mx} \Big\}, \\
\bar{\omega}\left({\bm k}\right)&\equiv\left[1+\left(\frac{2F^2}{v\Sigma}\right)^2\left(\bm{k}^2+\frac{D_3}{F^2}\left(\alpha _1 -\alpha _2\right)\right)\right]^{1/2}, \notag  \\
\omega^{II}&\equiv \mu H+\frac{v^2\Sigma}{2F^2}\left(-1+\bar{\omega}\right) \nonumber \\
&\simeq \mu H+\frac{\alpha _1 -\alpha _2}{\Sigma}D_3+\frac{F^2}{\Sigma}\bm{k}^2 , \notag \\
\omega^{M}&\equiv -\mu H+\frac{v^2\Sigma}{2F^2}\left(1+\bar{\omega}\right) \nonumber \\
&\simeq \frac{\Sigma}{F_t^2}-\mu H
+\frac{\alpha _1 -\alpha _2}{\Sigma}D_3
+\frac{F^2}{\Sigma}\bm{k}^2 .
\end{align}
The amplitudes for the scattering processes, $\pi^{II}({\bm k_1})+\pi^{II}({\bm k_2}) \rightarrow \pi^{II}({\bm k_3})+\pi^{II}({\bm k_4}),\pi^{II}({\bm k_1})+\pi^{M}({\bm k_2}) \rightarrow \pi^{II}({\bm k_3})+\pi^M({\bm k_4}),$ and $\pi^{M}({\bm k_1})+\pi^{M}({\bm k_2}) \rightarrow \pi^{M}({\bm k_3})+\pi^M({\bm k_4})$, are given by
\begin{align}
&M\left[\pi^{II}({\bm k_1})+\pi^{II}({\bm k_2}) \rightarrow \pi^{II}({\bm k_3})+\pi^{II}({\bm k_4})\right] \notag \\
&=\frac{1}{\sqrt{\bar{\omega } _1\bar{\omega}  _2 \bar{\omega}  _3 \bar{\omega } _4}}\Bigg[-\frac{1}{3\Sigma}\left(\omega_1^{II}+\omega_2^{II}\right) \nonumber \\
&\quad +\frac{F^2}{3\Sigma ^2}\left(4\bm{k}_1\cdot\bm{k}_2 +2\bm{k}_3\cdot\bm{k}_4 +\bm{k}_1^2+\bm{k}_2^2 \right) \notag \\
&\quad +\frac{2}{3\Sigma}\mu H+\frac{16}{3}\left(\alpha _1 -\alpha _2 \right) \frac{D_3}{\Sigma^2}\Bigg] \notag \\
&\simeq \frac{1}{\sqrt{\bar{\omega } _1\bar{\omega}  _2 \bar{\omega}  _3 \bar{\omega } _4}}\Biggl[\frac{2F^2}{\Sigma^2}\left({\bm k}_1 \cdot {\bm k}_2 \right) \nonumber \\
&\quad +\frac{14}{3}\left(\alpha_1-\alpha_2\right)\frac{D_3}{\Sigma^2}+O(\bm{k_i}^4,H^2,D^2)\Biggr] ,\\
&M\left[\pi^{II}({\bm k_1})+\pi^{M}({\bm k_2}) \rightarrow \pi^{II}({\bm k_3})+\pi^{M}({\bm k_4})\right] \notag \\
&=\frac{1}{\sqrt{\bar{\omega } _1\bar{\omega}  _2 \bar{\omega}  _3 \bar{\omega } _4}}\Bigg[\frac{1}{6\Sigma}\left(\omega_2^M+\omega_4^M-\omega_1^{II}-\omega_3^{II}\right) \nonumber \\
&\quad -\frac{F^2}{3\Sigma ^2}\left(4\bm{k}_2\cdot\bm{k}_3 +2\bm{k}_1\cdot\bm{k}_4- \bm{k}_2^2-\bm{k}_3^2  \right) \notag \\
&\quad +\frac{2}{3\Sigma}\mu H+\frac{16}{3}\left(\alpha _1 -\alpha _2 \right) \frac{D_3}{\Sigma^2}\Bigg]\notag \\
&\simeq \frac{1}{\sqrt{\bar{\omega } _1\bar{\omega}  _2 \bar{\omega}  _3 \bar{\omega } _4}}\Bigg[\frac{1}{3F_t^2}\nonumber \\
&\quad +\frac{F^2}{6\Sigma ^2}\left(-\bm{k}_1^2+3\bm{k}_2^2+\bm{k}_3^2+\bm{k}_4^2-4\bm{k}_1\cdot\bm{k}_4-8\bm{k}_2\cdot\bm{k}_4\right) \notag \\
&\quad +\frac{16}{3}\left(\alpha _1 -\alpha _2\right) \frac{D_3}{\Sigma ^2}\Bigg] , \\
&M\left[\pi^{M}({\bm k_1})+\pi^{M}({\bm k_2}) \rightarrow \pi^{M}({\bm k_3})+\pi^{M}({\bm k_4})\right] \notag \\
&=\frac{1}{\sqrt{\bar{\omega } _1\bar{\omega}  _2 \bar{\omega}  _3 \bar{\omega } _4}}\Bigg[\frac{1}{3\Sigma}\left(\omega_1^M+\omega_2^M\right)\nonumber \\
&\quad +\frac{F^2}{3\Sigma ^2}\left(4\bm{k}_1\cdot\bm{k}_2 +2\bm{k}_3\cdot\bm{k}_4 +\bm{k}_1^2+\bm{k}_2^2\right) \notag \\
&\quad +\frac{2}{3\Sigma}\mu H+\frac{16}{3}\left(\alpha _1 -\alpha _2 \right) \frac{D_3}{\Sigma^2}\Bigg], \notag \\
&\simeq \frac{1}{\sqrt{\bar{\omega } _1\bar{\omega}  _2 \bar{\omega}  _3 \bar{\omega } _4}}\Biggl[\frac{2}{3F_t^2}+\frac{2F^2}{3\Sigma ^2}\left(\left(\bm{k}_1+\bm{k}_2\right)^2+\bm{k}_1\cdot \bm{k}_2\right)
\nonumber \\
&\quad +6\left(\alpha _1 -\alpha _2\right) \frac{D_3}{\Sigma ^2}\Biggr], 
\end{align}
with $\bar{\omega}_i \equiv \bar{\omega}\left(\bm{k}_i\right)$. 
Note that the scattering amplitudes do not depend on the magnetic field but 
 on the anisotropic effect. 
Using the definition given in Appendix \ref{scattering length}, the scattering lengths are given by
\begin{align}
&a^{II+II\rightarrow II+II} \nonumber \\
&=\frac{\Sigma}{4F^2\bar{\omega}\left(0\right)}\left[-\frac{v^2}{3F^2}\left(\bar{\omega}(0)-1\right)+\frac{16}{3}\left(\alpha_1-\alpha_2\right)\frac{D_3}{\Sigma^2}\right] \notag \\
&\simeq \frac{7}{6F^2}\left(\alpha_1-\alpha_2\right)\frac{D_3}{\Sigma},\\
&a^{II+M\rightarrow II+M} \nonumber \\
&=\frac{\Sigma}{4F^2\bar{\omega}\left(0\right)}\left[\frac{1}{3F_t^2}+\frac{16}
{3}\left(\alpha_1-\alpha_2\right)\frac{D_3}{\Sigma^2}\right] \nonumber \\
&\simeq\frac{\Sigma}{4F^2}\left[\frac{1}{3F_t^2}+\frac{16}
{3}\left(\alpha_1-\alpha_2\right)\frac{D_3}{\Sigma^2}\right], \\
&a^{M+M\rightarrow M+M} \nonumber \\
&=\frac{\Sigma}{4F^2\bar{\omega}\left(0\right)}\left[\frac{v^2}{3F^2}\left(\bar{\omega}(0)+1\right)+\frac{16}{3}\left(\alpha_1-\alpha_2\right)\frac{D_3}{\Sigma^2}\right]\notag \\
&\simeq \frac{\Sigma}{4F^2}\left[\frac{2}{3F_t^2}+6\left(\alpha_1-\alpha_2\right)\frac{D_3}{\Sigma^2}\right] .
\end{align}
Again, all the scattering lengths are finite and only depend on the anisotropy.

The existence of the nonzero magnon scattering length has an important implication in few-body systems.
The nonzero scattering length is induced by the four-magnon contact vertex without derivatives
in Eq.~\eqref{L4pi}, because the scattering length is defined at zero momentum. 
The above results are obtained by the perturbative calculation in the counting scheme
~\eqref{counting} where the strength of the external field is counted as $\mathcal{O}(p)$.
As a consequence, the scattering length obtained should not be very large. 

On the other hand, we can modify the power counting scheme in the presence of the contact 
interaction. Considering the small momentum limit with a fixed strength of the external field,
we may regard the external field as $\mathcal{O}(1)$. This is analogous to the chiral EFT for
the nuclear force, where the leading order term contains the $\mathcal{O}(1)$ contact interaction. 
In this case, systematic perturbation theory to describe the two-nucleon system with a large 
scattering length can be formulated, by the nonperturbative resummation of the $\mathcal{O}(1)$ 
contact term \cite{Kaplan:1998tg,Kaplan:1998we}. In general, nonrelativistic EFT with a contact 
interaction has two fixed points, corresponding to the vanishing scattering length (non-interacting
 limit) and the infinitely large scattering length (unitary limit) \cite{Braaten:2004rn,Braaten:2007nq}.
While the original counting scheme~\eqref{counting} corresponds to the expansion around the non-interacting 
limit, the new counting scheme gives the expansion around the unitary limit \cite{Kaplan:1998tg,Kaplan:1998we}. 
We thus conclude that the EFT in this work can describe the magnon systems with a large scattering length, 
with an appropriately modified power counting scheme.

When the two-body scattering length is infinitely large, the three-body system is known to exhibit
the Efimov effect \cite{Braaten:2004rn}. In fact, the Efimov effect for magnons is shown to be induced 
by the anisotropic effect \cite{Nishida:2012hf,nishida2013electron}. This is consistent with
the presence of the contact term in the EFT with the anisotropic effect.

Moreover, we note that the scattering length of the gapped mode in the ferrimagnet remains finite 
without the anisotropic effect as in Eq.~\eqref{scatteringlengthMM}. In this case, although the contact 
term does not explicitly appear in the Lagrangian, the time derivative term generates an effective contact
interaction which is proportional to the gap energy.\footnote{Strictly speaking, this kind of effective 
contact term also contributes to the scattering length with the anisotropic effect.} We expect that the 
Efimov effect for magnons in a ferrimagnet can be realized even in the absence of the anisotropic effect.

\section{Summary and concluding remarks}

In this paper, we have discussed the low-energy effective field theory 
for spin systems including a ferrimagnet, 
and the scattering processes of the magnons 
as the Nambu-Goldstone(NG) modes. 
On the basis of the Lagrangians, 
we have obtained the dispersion relations of the magnons as the NG modes, 
which coincide with the microscopic results. 
In particular the case of the ferrimagnet is worth mentioning, where 
the NG modes have two-types of dispersion relations, 
$\omega \propto \bm{k}^2,\, \bm{k}^2+ m^2$, i.e., a massless mode with a quadratic momentum dependence
and a gapped mode, both of
which come from the spontaneous symmetry breaking. 
We have shown that the Lagrangian 
including only terms with one time derivative describes 
the system with just one order parameter 
given by the commutation relation between the broken charge and the charge density (magnetization), 
while the Lagrangian including both terms with one and two time derivatives describes the system 
with two order parameters (magnetization and staggered magnetization). 
This result also supports the good correspondence between the EFTs and the 
microscopic theories of the spin systems.
Furthermore, we have established the power counting scheme for the Lagrangian which confirmed the validity of the gapped mode, and clarified the systematic expansion.

To determine the particle (magnon) states and discuss the scattering process, 
we have derived the Hamiltonian from the Lagrangian. In the ferromagnet case, 
where no term with two time derivatives
exists in the Lagrangian, 
the Hamiltonian has been constructed using the Dirac-Bergmann method for constraint systems, 
while in the antiferromagnet  and the ferrimagnet cases the Hamiltonian has been constructed 
by the conventional method. Furthermore,
the quantization of the system has been done by equating the commutation relation with
the Dirac brackets 
in the ferromagnet and the Poisson bracket in the antiferromagnet and the ferrimagnet cases.
Thus we have found that
 one magnon state as the NG mode appears in the ferromagnet, 
while  two magnon states do so
in the antiferromagnet and the ferrimagnet.

Then we have calculated the scattering amplitudes for magnon-magnon scattering processes 
in the nonrelativistic notation. For the magnon-magnon scattering in
 the ferromagnet and antiferromagnet, 
we have reproduced the results given in a previous study \cite{Hofmann:1998pp}. 
In addition, we have obtained the amplitudes for the three types 
of the magnon-magnon scattering processes in the ferrimagnet. 
Remarkably enough, the scattering lengths of the processes involving
  the gapped state are finite even without explicit symmetry
  breaking. The strengths of the scattering lengths are related
  to the gap of the mode, because both originate in the spontaneous
  symmetry breaking. The relation of the scattering length and
  the gap will be useful to identify the gapped NG mode in real
  materials. The finite scattering length also implies that the
  Efimov effects come into play in the ferrimagnet even without
  the external fields, in contrast to the ferromagnet discussed
  in Refs.\cite{Nishida:2012hf,nishida2013electron}.

Finally, we have discussed the effects
 of the magnetic fields and the anisotropy as explicit symmetry breaking effects. 
The effective Lagrangian and the physical states of magnons have been constructed, 
and the scattering amplitudes have been calculated. 
All the amplitudes for the magnon-magnon scattering processes 
are affected by the anisotropy but not by the magnetic field.
The irrelevance of the magnetic field may 
be attributed to the identification 
of the magnetic field with the SO(3) gauge-field. 
In a recent cold atom experiment, the dispersion relation of the
  magnon in the ferromagnetic phase of spinor Bose-Einstein condensate
  was successfully measured \cite{marti2014coherent}. In addition,
  realization of the ferrimagnetic state is also possible, for
  instance, using Kagome lattices \cite{chen2012kondo, yamada2011mott}.
  These developments, together with the Efimov effect, will be useful
  to experimentally realize the findings in this paper.

\section*{Acknowledgments}
The authors are grateful to Yoshimasa Hidaka for suggesting that we examine the relation between the order parameter and the coefficients from the perspective of the EFT as discussed in Sec.\ref{condensates_and_coefficients}. The authors thank Yoshimasa Hidaka, Tomoya Hayata, Satoshi Fujimoto, and Sinya Aoki for useful discussions and comments. S.~G.~ is supported by JSPS KAKENHI Grants No. 25287046, SPIRE (Strategic Program for
Innovative REsearch) and  a grant from La R\'egion Centre (France). Y.~K.~is supported by a Grant-in-Aid for Scientific Research from the JSPS Fellows (No.15J01626).
 T.~H.~is supported by JSPS KAKENHI Grants No. 24740152.
T.~K.~is supported by JSPS KAKENHI Grants No. 24340054.
  T.~H.~ and T.~K.~ are supported by the Yukawa International Program for Quark-Hadron Sciences (YIPQS).

\appendix

\section{Scattering length, phase shift, and scattering amplitude in the nonrelativistic notation}
\label{scattering length}
In the nonrelativistic notation, the normalization of the states is defined by
\begin{align}
\left< f |i \right> = \delta_{fi} .
\end{align}

The $S$ matrix element $S_{fi}$ and the scattering amplitude $M_{fi}$ for the scattering process $i \rightarrow f$ are related as follows:
\begin{align}
S_{fi}=\delta_{fi} - i \left(2\pi\right)^4 \delta^{4}\left(P_f - P_i \right)M_{fi},
\end{align}
where $P_i$ and $P_f$ denote the total momentum of the initial and final states, respectively. 
The unitary condition, $SS^\dagger =1$, is reduced to 
\begin{align}
-2\mathrm{Im}M_{fi} = \sum _n \left(2\pi \right)^4 M_{fn} M_{ni}^\dagger \delta ^{4}\left(P_n -P_i \right).
\end{align}
For two-particle scattering processes below inelastic thresholds 
in the center of mass frame, the summation is reduced to the form given in terms of
energies $E_k,\,E_i$ as,
\begin{align}
&\sum _n \delta^4\left(P_n-P_i \right)\nonumber \\
&\rightarrow \int \frac{d^3k_1}{\left(2\pi \right)^3}\int 
\frac{d^3k_2}{\left(2\pi \right)^3}\delta^3({\bm k}_1+{\bm k}_2)\delta\left(E_k-E_i\right) ,
\end{align}
 and the scattering amplitude 
is found to be written as $M_{fi}=M\left({\bm q_f},{\bm q_i}\right)$ 
with the tree-dimensional initial and final momenta, ${\bm q_i}, {\bm q_f}$.
The unitary condition of the scattering amplitude is rewritten as
\begin{align}
&-2\mathrm{Im}M\left({\bm q_f},{\bm q_i}\right) \nonumber \\
&= \int \frac{d^3k}{\left(2\pi\right)^2} M\left({\bm q_f},{\bm k}\right) M^\dagger\left({\bm k},{\bm q_i}\right) \delta \left(E\left( \left|{\bm k}\right|\right) -E_i \right).
\end{align}
Using the spherical harmonic functions $Y_{lm}$ and the Legendre polynomials $P_l$, the partial wave amplitude $M^l_{fi}$ is given by 
\begin{align}
M ({\bm q_f}, {\bm q_i})&=\sum _l \left(2l+1\right)P_l \left(\cos \theta\right)M_l\left(|{\bm q_f}|,|{\bm q_i}| \right)  \notag \\
&= 4\pi\sum_{l,m} Y_{lm}\left(\Omega _f\right) Y^\dagger _{lm} \left(\Omega _i \right)M_l\left(|{\bm q_f}|,|{\bm q_i}|\right),
\end{align}
with scattering angle $\theta$ and solid angles $\Omega_f,\, \Omega_i$ for the final and initial state. 
By performing the partial wave expansion and the angular integral, the unitary condition 
is reduced to
\begin{align}
&-2\sum_{l} \left(2l+1\right)P_l\left(\cos\theta\right)\mathrm{Im}M_l ({\bm q_f}, {\bm q_i}) \nonumber \\
&=\int^{\infty}_{0} \left|{\bm k}\right| ^2d\left|{\bm k}\right|\sum _{l,m} Y_{lm}\left(\Omega_f \right)Y_{lm}^\dagger \left(\Omega _i \right)  \notag \\
&\quad \times M_l\left(|{\bm q_f}|,|{\bm k}|\right)M_l^\dagger \left(|{\bm k}|,|{\bm q_i}|\right)\delta\left(E\left( \left|{\bm k}\right|\right)-E_i\right)\notag \\
&=\sum_{l}\frac{2l+1}{4\pi}P_l \left( \cos \theta \right)\int^{\infty}_{0} \left|{\bm k}\right| ^2d\left|{\bm k}\right|\notag \\
&\quad \times M_l\left(|{\bm q_f}|,|{\bm k}|\right)M_l^\dagger \left(|{\bm k}|,|{\bm q_i}|\right)\delta\left( E\left( \left|{\bm k}\right|\right)-E_i\right).
\end{align}
Thus the unitary condition is given in terms of the partial wave amplitude,
\begin{align}
&\quad -2\mathrm{Im}M_l (q,q) \nonumber \\
&=\frac{1}{4\pi}\int^{\infty}_{0} \left|{\bm k}\right| ^2d\left|{\bm k}\right|M_l\left(q,|{\bm k}|\right)M_l^\dagger \left(|{\bm k}|,q\right)\delta\left( E\left( \left|{\bm k}\right|\right)-E_i\right) \notag \\
&=\frac{1}{4\pi}\left|  \frac{\partial E\left(  q\right)} {\partial q}\right|^{-1}q^2 M_l\left(q,q\right)M_l^\dagger \left( q,q\right),
\end{align} 
where we used $q\equiv |{\bm q_f}|=|{\bm q_i}|$. Using the unitary condition, the phase shift $\delta _l$ is defined through the partial wave amplitude $M_l$: 
\begin{align}
M_l (q) =\frac{8\pi}{q^2}\left|  \frac{\partial E\left( q\right)} {\partial q}\right|\sin \delta _l e^{i\delta _l}.
\end{align}
Using this expression, the scattering length is defined from the amplitude:
\begin{align}
a= \lim_{q\rightarrow 0}q\left|  \frac{\partial E\left( q\right)} {\partial q}\right|^{-1}M(q,\theta),
\end{align} 
with the tree-dimensional momenta $q$ and the scattering angle $\theta$ in the center of mass. The definition is opposite in sign to the standard definition $a_{0}$ with the effective range expansion $k\cot\delta_{0}=-1/a_{0}+\mathcal{O}(k^{2})$.


\bibliographystyle{ptephy}
\bibliography{spinwave}

\begin{thebibliography}{10}

\bibitem{Weinberg:1978kz}
Steven Weinberg, Physica, {\bf A96}, 327 (1979).

\bibitem{Gasser:1983yg}
J.~Gasser and H.~Leutwyler, Annals Phys., {\bf 158}, 142 (1984).

\bibitem{Gasser:1984gg}
J.~Gasser and H.~Leutwyler, Nucl. Phys., {\bf B250}, 465 (1985).

\bibitem{Leutwyler:1993gf}
H.~Leutwyler, Phys. Rev., {\bf D49}, 3033--3043 (1994),
  {{arXiv:hep-ph/9311264}}.

\bibitem{Burgess:1998ku}
C.~P. Burgess, Phys. Rept., {\bf 330}, 193--261 (2000),
  {{arXiv:hep-th/9808176}}.

\bibitem{Brauner:2010wm}
Tomas Brauner, Symmetry, {\bf 2}, 609--657 (2010),  {{arXiv:1001.5212}}.

\bibitem{Scherer:2012xha}
Stefan Scherer and Matthias~R. Schindler, Lect. Notes Phys., {\bf 830},
  pp.1--338 (2012).

\bibitem{Lee:2015bva}
Tong-Gyu Lee, Eiji Nakano, Yasuhiko Tsue, Toshitaka Tatsumi, and Bengt Friman,
  Phys. Rev., {\bf D92}(3), 034024 (2015),  {{arXiv:1504.03185}}.

\bibitem{Hidaka:2015xza}
Yoshimasa Hidaka, Kazuhiko Kamikado, Takuya Kanazawa, and Toshifumi Noumi,
  Phys. Rev., {\bf D92}(3), 034003 (2015),  {{arXiv:1505.00848}}.

\bibitem{Hasenfratz:1993vf}
P.~Hasenfratz and F.~Niedermayer, Z. Phys., {\bf B92}, 91 (1993),
  {{arXiv:hep-lat/9212022}}.

\bibitem{Leutwyler:1996er}
H.~Leutwyler, Helv. Phys. Acta, {\bf 70}, 275--286 (1997),
  {{arXiv:hep-ph/9609466}}.

\bibitem{roman9709298effective}
J.~M. Roman and J.~Soto, Int. J. Mod. Phys. B, {\bf 13}, 755 (1999),
  {{cond-mat/9709298}}.

\bibitem{Hofmann:1998pp}
Christoph~P. Hofmann, Phys. Rev., {\bf B60}, 388 (1999),
  {{arXiv:cond-mat/9805277}}.

\bibitem{Hofmann:2001ck}
Christoph~P. Hofmann, Phys. Rev., {\bf B65}, 094430 (2002),
  {{arXiv:cond-mat/0106492}}.

\bibitem{Nishida:2012hf}
Yusuke Nishida, Yasuyuki Kato, and Cristian~D. Batista, Nature Phys., {\bf 9},
  93--97 (2013),  {{arXiv:1208.6214}}.

\bibitem{nishida2013electron}
Yusuke Nishida, Physical Review B, {\bf 88}(22), 224402 (2013).

\bibitem{Hyodo:2013zxa}
Tetsuo Hyodo, Tetsuo Hatsuda, and Yusuke Nishida, Phys. Rev., {\bf C89}(3),
  032201 (2014),  {{arXiv:1311.6289}}.

\bibitem{Nielsen:1975hm}
Holger~Bech Nielsen and S.~Chadha, Nucl. Phys., {\bf B105}, 445 (1976).

\bibitem{Miransky:2001tw}
V.~A. Miransky and I.~A. Shovkovy, Phys. Rev. Lett., {\bf 88}, 111601 (2002),
  {{arXiv:hep-ph/0108178}}.

\bibitem{Schafer:2001bq}
Thomas Schafer, D.~T. Son, Misha~A. Stephanov, D.~Toublan, and J.~J.~M.
  Verbaarschot, Phys. Lett., {\bf B522}, 67--75 (2001),
  {{arXiv:hep-ph/0108210}}.

\bibitem{Nambu:2004yia}
Yoichiro Nambu, J. Statist. Phys., {\bf 115}(1/2), 7--17 (2004).

\bibitem{Hidaka:2012ym}
Yoshimasa Hidaka, Phys. Rev. Lett., {\bf 110}(9), 091601 (2013),
  {{arXiv:1203.1494}}.

\bibitem{Watanabe:2012hr}
Haruki Watanabe and Hitoshi Murayama, Phys. Rev. Lett., {\bf 108}, 251602
  (2012),  {{arXiv:1203.0609}}.

\bibitem{Kapustin:2012cr}
Anton Kapustin (2012),  {{arXiv:1207.0457}}.

\bibitem{Nicolis:2012vf}
Alberto Nicolis and Federico Piazza, Phys. Rev. Lett., {\bf 110}(1), 011602,
  [Addendum: Phys. Rev. Lett.110,039901(2013)] (2013),  {{arXiv:1204.1570}}.

\bibitem{Gongyo:2014sra}
Shinya Gongyo and Shintaro Karasawa, Phys. Rev., {\bf D90}(8), 085014 (2014),
  {{arXiv:1404.1892}}.

\bibitem{Hayata:2014yga}
Tomoya Hayata and Yoshimasa Hidaka, Phys. Rev., {\bf D91}, 056006 (2015),
  {{arXiv:1406.6271}}.

\bibitem{Beekman:2014cba}
Aron~J. Beekman, Annals Phys., {\bf 361}, 461--489 (2015),
  {{arXiv:1408.1691}}.

\bibitem{Andersen:2014ywa}
Jens~O. Andersen, Tomas Brauner, Christoph~P. Hofmann, and Aleksi Vuorinen,
  JHEP, {\bf 08}, 088 (2014),  {{arXiv:1406.3439}}.

\bibitem{Watanabe:2014fva}
Haruki Watanabe and Hitoshi Murayama, Phys. Rev., {\bf X4}(3), 031057 (2014),
  {{arXiv:1402.7066}}.

\bibitem{Coleman:1969sm}
Sidney~R. Coleman, J.~Wess, and Bruno Zumino, Phys. Rev., {\bf 177}, 2239--2247
  (1969).

\bibitem{Callan:1969sn}
Curtis~G. Callan, Jr., Sidney~R. Coleman, J.~Wess, and Bruno Zumino, Phys.
  Rev., {\bf 177}, 2247--2250 (1969).

\bibitem{kittel1963quantum}
Charles Kittel and Ching-yao Fong,
\newblock {\em Quantum theory of solids}, volume~33,
\newblock  (Wiley New York, 1963).

\bibitem{brehmer1997low}
S~Brehmer, HJ~Mikeska, and Shoji Yamamoto, Journal of Physics: Condensed
  Matter, {\bf 9}(19), 3921 (1997).

\bibitem{pati1997density}
Swapan~K Pati, S~Ramasesha, and Diptiman Sen, Journal of Physics: Condensed
  Matter, {\bf 9}(41), 8707 (1997).

\bibitem{Watanabe:2013uya}
Haruki Watanabe, Tomas Brauner, and Hitoshi Murayama, Phys. Rev. Lett., {\bf
  111}(2), 021601 (2013),  {{arXiv:1303.1527}}.

\bibitem{landau1960classical}
LD~Landau and EM~Lifshitz,
\newblock {\em Classical mechanics},
\newblock  (Pergamon Press, Oxford, 1960).

\bibitem{Dirac:1950pj}
Paul A.~M. Dirac, Can. J. Math., {\bf 2}, 129--148 (1950).

\bibitem{Bergmann:1949zz}
Peter~G. Bergmann, Phys. Rev., {\bf 75}, 680--685 (1949).

\bibitem{Dyson:1956zz}
Freeman~J. Dyson, Phys. Rev., {\bf 102}, 1230--1244 (1956).

\bibitem{Weinberg:1966kf}
Steven Weinberg, Phys. Rev. Lett., {\bf 17}, 616--621 (1966).

\bibitem{Kaplan:1998tg}
David~B. Kaplan, Martin~J. Savage, and Mark~B. Wise, Phys. Lett., {\bf B424},
  390--396 (1998),  {{arXiv:nucl-th/9801034}}.

\bibitem{Kaplan:1998we}
David~B. Kaplan, Martin~J. Savage, and Mark~B. Wise, Nucl. Phys., {\bf B534},
  329--355 (1998),  {{arXiv:nucl-th/9802075}}.

\bibitem{Braaten:2004rn}
Eric Braaten and H.~W. Hammer, Phys. Rept., {\bf 428}, 259--390 (2006),
  {{arXiv:cond-mat/0410417}}.

\bibitem{Braaten:2007nq}
Eric Braaten, Masaoki Kusunoki, and Dongqing Zhang, Annals Phys., {\bf 323},
  1770--1815 (2008),  {{arXiv:0709.0499}}.

\bibitem{marti2014coherent}
G~Edward Marti, Andrew MacRae, Ryan Olf, Sean Lourette, Fang Fang, and Dan~M
  Stamper-Kurn, Phys. Rev. Lett., {\bf 113}(15), 155302 (2014),
  {{arXiv:1404.5631}}.

\bibitem{chen2012kondo}
Yao-Hua Chen, Hong-Shuai Tao, Dao-Xin Yao, and Wu-Ming Liu, Phys. Rev. Lett.,
  {\bf 108}(24), 246402 (2012),  {{arXiv:1201.0654}}.

\bibitem{yamada2011mott}
Atsushi Yamada, Kazuhiko Seki, Robert Eder, and Yukinori Ohta, Phys. Rev., {\bf
  B83}(19), 195127 (),  {{arXiv:1101.3645}}.

\end{thebibliography}

%

\end{document}